\documentclass[a4paper,11pt]{article}

\usepackage[english]{babel}

\usepackage[latin1]{inputenc}
\usepackage[toc,page]{appendix}
\usepackage{amsfonts,amsbsy,bm,euscript,mathrsfs}
\usepackage{amssymb,stmaryrd,faktor,slashed}
\usepackage{color}

\usepackage[nosort]{cite}

\usepackage{authblk, physics, multirow, comment}

\usepackage{graphicx}
\usepackage[verbose]{wrapfig}
\usepackage{caption}
\usepackage{float}

\numberwithin{equation}{section}

\usepackage{enumitem}

\newcommand{\ie}{{\it i.e.\ }}

\newcommand{\vol}{\mathrm{vol}}

\renewcommand{\dd}{\mathrm{d}}
\newcommand{\ee}{\mathrm{e}}

\newcommand{\bbR}{\mathbb{R}}

\newcommand{\Gs}[1]{\Gamma(#1)}

\newenvironment{thm}
{\par\medskip\noindent\textbf{Theorem.}\itshape}
{\par\medskip}

\usepackage{jheppub}

\arxivnumber{2603.24534}

\title{\boldmath \Large\bf \Large\bf $Spin(n,n)\times\mathbb{R}^+$ Generalised Geometry and Consistent Truncations on Branes}

\author[a]{Jieming Lin\,,}
\author[a]{K. S. Stelle}
\author[a]{and Daniel Waldram}
\affiliation[a]{Abdus Salam Centre for Theoretical Physics, Imperial College London,\\ Prince Consort Road, London, SW7 2AZ, UK}

\emailAdd{jieming.lin22@imperial.ac.uk, d.waldram@imperial.ac.uk}

\abstract{In this note we show how the consistent truncations on half-supersymmetric branes of Leung and Stelle and Lin, Skrzypek and Stelle fit into the general exceptional generalised geometry analysis of Cassani \emph{et al.}. Each solution defines a torsion-free $Spin(n)$ structure in the $Spin(n,n)\times \mathbb{R}^+$ generalised geometry introduced by Strickland--Constable, where $n$ is the dimension of the space transverse to the brane. Embedding this into the appropriate exceptional generalised geometry then defines the truncation. As a by-product we derive a new consistent truncation on the IIA NS5-brane to six-dimensional $\mathcal{N}=(2,0)$ supergravity coupled to a tensor multiplet, and new consistent truncations on the D6- and D7-branes to seven- and eight-dimensional pure half-maximal supergravity respectively.  }

\begin{document}

\begin{flushright}\small{Imperial--TP--2026--DW--1}\end{flushright}
\vspace{0.1cm}

\maketitle
\flushbottom

\section{Introduction}

Consistent truncations are a standard tool in gravitational theories that can provide a relatively easy way of finding new solutions. Dimensionally reducing the theory on a manifold $M$, the key task is to find a finite subset of the infinite tower of Kaluza--Klein modes on $M$ such that every solution of the lower-dimensional theory uplifts, via the truncation ansatz, to a solution of the higher-dimensional one. Generically such consistent truncations are relatively rare and hard to construct~\cite{Duff:1984hn, Cveti__2000, Lu:2006dh}.

The simplest consistent truncations arise from taking $M$ to be a Lie group (or a quotient thereof). For such Scherk--Schwarz reductions consistency follows from the group structure~\cite{Scherk:1978ta,Scherk:1979zr}. There are in addition the famous maximally supersymmetric sphere reductions of eleven-dimensional ~\cite{deWit:1986oxb,Nastase:1999cb,Nastase:1999kf} and type IIB~\cite{Cvetic:2000nc,Lee:2014mla,Baguet:2015sma} supergravities and multiple  examples arising from reductions on backgrounds preserving some fraction of supersymmetry (for early work in this direction see~\cite{Gauntlett:2007ma,Kashani-Poor:2007nby,Gauntlett:2007sm,Gauntlett:2009zw,Cassani:2009ck,Cassani:2010uw,Liu:2010sa,Gauntlett:2010vu,Skenderis:2010vz}). Experience with these examples, and the relation to AdS-CFT, leads to a natural conjecture~\cite{Duff:1984hn,Pope:1987ad,Gauntlett:2007ma} that can be stated in different ways but reduces to the claim that whenever one has a solution of the higher-dimensional theory preserving $n$ supersymmetries, to either flat space or anti-de Sitter space, there should be a consistent truncation to pure supergravity with $n$ supercharges. Perhaps the simplest example arises from the truncation to pure supergravity of the consistent truncation to the universal sector on a nearly-K\"ahler or Calabi--Yau manifold given in~\cite{Terrisse:2019usq,Tsimpis:2020ysl}, with the fermionic sector in the latter case given explicitly in~\cite{Lin:2024eqq}. Of particular relevance to this note, the conjecture was also explicitly demonstrated to be true for several half-supersymmetric $p$-brane backgrounds in~\cite{Leung:2022nhy}, and subsequently generalised  to a systematic construction for general half-supersymmetric $p$-brane backgrounds in~\cite{Lin:2025ucv}, using an ansatz based on a special class of intersecting brane solutions. 

Remarkably, all known consistent truncations can be systematically understood~\cite{Cassani:2019vcl} using $G$-structures and their extension to exceptional generalised geometry~\cite{Lee:2014mla, Hohm:2014qga, Baguet:2015sma, Lee:2015xga, Malek:2015hma, Baguet:2015iou, Ciceri:2016dmd,Cassani:2016ncu, Malek:2016bpu, duBosque:2017dfc, Malek:2017njj, Inverso:2017lrz, Malek:2017cle, Malek:2018zcz, Malek:2019ucd, Cassani:2019vcl, Samtleben:2019zrh, Cassani:2020cod, Malek:2020jsa, Josse:2021put, Bugden:2021wxg, Bugden:2021nwl, Galli:2022idq, Bossard:2022wvi, Butter:2022iza, Hulik:2023aks, Bossard:2023jid, Josse:2023lsp, Hassler:2023axp, Blair:2024ofc, Inverso:2024xok, Guarino:2024gke, Duboeuf:2024tbd, Hassler:2025qhh, Rovere:2025jks, Pico:2025cmc, Josse:2025uro, Pico:2026rji}. First proposed in \cite{Hull:2007zu,PiresPacheco:2008qik} and then fully developed in~\cite{Coimbra:2011ky, Coimbra:2012af}, exceptional generalised geometry is a reformulation of eleven-dimensional and type II supergravity using an extension of the conventional tangent bundle to include the form-field gauge symmetries of supergravity. Strikingly the structure group extends the exceptional group $E_{k(k)}$. In its original form it captures the full Kaluza--Klein spectrum of the scalar and fermionic degrees of freedom of any compactification. The full extension to vector and tensor modes was done in~\cite{Hohm:2013pua,Hohm:2013vpa,Hohm:2013uia}. One finds~\cite{Cassani:2019vcl} that all known consistent truncations arise from a generalised $G$-structure with $G\subset E_{k(k)}$, that is a geometry where the structure group reduces to $G$, provided the structure has the geometrical property of having constant singlet intrinsic torsion. Under these assumptions one can prove that an ansatz that keeps all modes that are singlets under $G$ leads to a consistent truncation~\cite{Cassani:2019vcl}. Furthermore the field content and gauging of the reduced theory is completely determined by the structure and the intrinsic torsion and the construction allows one to prove the pure supergravity conjecture mentioned above.   

In this note, we revisit the $p$-brane truncations of~\cite{Leung:2022nhy,Lin:2025ucv} from the perspective of exceptional generalised geometry showing that they indeed admit a uniform description within the generalised $G$-structure framework of~\cite{Cassani:2019vcl}. We find that they all appear in the same way, via the embedding of a simpler ``$Spin(n,n)$ generalised geometry'', first introduced in~\cite{Strickland-Constable:2013xta}, inside the exceptional generalised geometry, where $n$ is the dimension of the space transverse to the brane. 
The $G$-structure defining the truncation is $Spin(n)_S$, corresponding to the first factor in the embedding of the maximally compact subgroup $Spin(n)_S\times Spin(n)\subset Spin(n,n)$. 
It is furthermore torsion-free, meaning the resulting truncated supergravities are ungauged. Under the embedding $Spin(n)_S\times Spin(n)\subseteq \tilde{H}_k$, where $\tilde{H}_k$ is the double-cover of the maximally compact subgroup $H_k\subset E_{k(k)}$, the $\tilde{H}_k$ representation that contains the supersymmetry parameters decomposes as $(s,\mathbf{1})\oplus (\mathbf{1},s)$ where $s$ is the spinor representation. The parameters in the second factor are singlets under $Spin(n)_S$ and hence the truncations are half-supersymmetric.
In addition to finding the relevant generalised $G$-structures, we also complete the analysis of~\cite{Leung:2022nhy,Lin:2025ucv} to include the type IIA NS5-brane and the D6- and D7-brane theories. The former is particularly interesting because unlike the other cases, the resulting supergravity contains matter, namely an additional tensor multiplet.

The remainder of this note is organised as follows. In Section~\ref{sec:gg}, after reviewing the construction of consistent truncations using $G$-structures in exceptional generalised geometry, we summarise the essential features of $Spin(n,n)\times\mathbb{R}^+$ generalised geometry relevant for consistent truncations, leaving the detailed construction to Appendix~\ref{app:sgg}. We also show that $p$-brane solutions admit $Spin(n)$-structures and construct the corresponding $Spin(n)$ singlets. In Section~\ref{sec:ct on brane}, we construct the explicit truncation ans\"atze for $p$-brane backgrounds with $3\leq p\leq 7$ using generalised geometry. We conclude and discuss possible future directions in Section~\ref{sec:conclusion}. The basic results on exceptional generalised geometry for M-theory and type IIB are summarised in Appendix~\ref{app:egg}. Finally, in Appendix~\ref{app:NS5A}, by way of a check, we prove the consistency of the truncation of type IIA supergravity to six-dimensional $(2,0)$ supergravity with one additional tensor multiplet using conventional techniques. 

\subsection*{Note added}

Kellogg S. Stelle passed away in October 2025, while the work in this note was being completed, and this paper has been prepared for publication by J.~Lin and D.~Waldram.

Kelly was a much-loved and highly valued colleague, friend and supervisor. He made multiple outstanding and long-lasting contributions to theoretical physics and to the theoretical physics community, and is deeply missed. (JL and DW.)

\section{$Spin(n,n)$ geometry and consistent truncations}\label{sec:gg}

\subsection{Consistent truncations from generalised $G$-structures}\label{sec:ct review}

We start by giving a brief overview of the formalism of generalised geometry and how it gives a systematic treatment of consistent truncations following~\cite{Cassani:2019vcl}. Consider type IIA, type IIB  or eleven-dimensional supergravity theory. The bosonic fields are tensors, that is representations of the $GL(10,\bbR)$ or $GL(11,\bbR)$ structure group of the spacetime. Consider now the restriction to spacetimes of the form $Z\times M$ where $Z$ is a $d$-dimensional Lorentzian space and $M$ is an $n$-dimensional ``internal'' space, which is not necessarily compact. Remarkably, if we drop the requirement of manifest $GL(10,\bbR)$ or $GL(11,\bbR)$ representations, the internal $GL(n,\bbR)$ structure group can be enhanced to the split exceptional group $E_{n(n)}$ in the case of eleven-dimensional supergravity, and to $E_{n+1(n+1)}$ in the case of the type II theories. In other words, the bosonic fields can be repacked as representations of $GL(d,\bbR)\times E_{n(n)}$ or $GL(d,\bbR)\times E_{n+1(n+1)}$. The fermions can be similarly repackaged into representations of $Spin(d-1,1)\times \tilde{H}_n$ or $Spin(d-1,1)\times \tilde{H}_{n+1}$ where $\tilde{H}_n$ is the double-cover of the maximally compact subgroup $H_n\subset E_{n(n)}$. It is important to emphasise that this is not a truncation, but simply a rewriting of the full ten- or eleven-dimensional theory. 

If we label the bosonic fields as scalars, vectors, two-forms etc depending on their $GL(d,\bbR)$ representation on $Z$ and write $x^\mu$ and $y^m$ for coordinates on $Z$ and $M$ respectively one finds
\begin{equation}
\label{eq:sugra-fields}
\begin{aligned}
   \text{scalars:} &&& & \hat{G}_{MN}(x,y) &\in \Gs{S^2 \hat{E}^*}\, ,\\[1mm]
   \text{vectors:} &&& & \hat{\mathcal{A}}_\mu{}^M(x,y) &\in \Gs{T^*Z\otimes \hat{E}}\, ,\\[1mm]
   \text{two-forms:} &&& & \hat{\mathcal{B}}_{\mu\nu}{}^{I}(x,y) &\in \Gs{\Lambda^2T^*Z\otimes \hat{N}}\, ,     
   \end{aligned}
\end{equation}
where $\hat{E}$ and $\hat{N}\subset S^2\hat{E}$ are particular vector bundles transforming in irreducible representations of $E_{n(n)}$ or $E_{n+1(n+1)}$, $S^2\hat{E}^*$ denotes the symmetric tensor product and the indices $M$ and $I$ denote the components of a vector in $\hat{E}$ and $\hat{N}$ respectively.\footnote{Note that $\hat{N}\subset S^2\hat{E}$ and \cite{Cassani:2019vcl} uses a pair $MN$ of indices in $S^2\hat{E}$ rather than a single index $I$ for the components of vector in $\hat{N}$.} Depending on the dimension of $Z$ one might have higher-form fields as well, but for the purposes of this paper the scalar, vector and two-form representations are sufficient. The object $\hat{G}_{MN}$ is known as the generalised metric and is invariant under the action of maximal subgroups $H_n$ or $H_{n+1}$, and $\hat{E}$ is known as the generalised vector bundle.

To be more concrete, consider the case of $n=5$ in eleven-dimensional supergravity.\footnote{Note that for simplicity, in this paragraph only, we violate the notational conventions summarised in Appendix~\ref{app:conventions}.} The bosonic fields in eleven dimensions are a metric $g$ and a three-form potential $A$, with a dual six-form potential $\tilde{A}$. We have the isomorphism $E_{5(5)}\simeq Spin(5,5)$ and the $\hat{E}$ and $\hat{N}$ bundles transform in the spinor $\mathbf{16^c}$ and vector $\mathbf{10}$ representations of $Spin(5,5)$ respectively. The bosonic fields arrange as 
\begin{equation}
\begin{aligned}
  \hat{E} & \simeq TM \oplus \Lambda^2T^*M \oplus \Lambda^5T^*M  &
  && \text{with} && &&
  \mathcal{A}_\mu{}^M &= ( g_\mu{}^m, A_{\mu mn}, \tilde{A}_{\mu m_1\dots m_5}) , \\
  \hat{N} & \simeq T^*M \oplus \Lambda^4T^*M &
  && \text{with} && &&
  \mathcal{B}_{\mu\nu}{}^{I} &= ( A_{\mu\nu m}, \tilde{A}_{\mu\nu m_1\dots m_4}) ,
\end{aligned}
\end{equation}
while the generalised metric is built from the internal components 
\begin{equation}
  \hat{G}_{MN} \sim ( g_{mn}, A_{mnp} ) . 
\end{equation}
and at each point $(x,y)\in Z\times M$ parameterises the coset $\mathbb{R}^+\times E_{5(5)} /H_5$ where $H_5\simeq Spin(5)\times Spin(5)$. 

This rewriting of supergravity is useful in many contexts, but of particular relevance here is that it leads to the main theorem of~\cite{Cassani:2019vcl} which states the following:
\begin{thm}
Let $k=n$ and $k=n+1$ for eleven-dimensional and type II supergravity respectively on $Z\times M$. If $M$ admits a generalised $G_S$-structure with $G_S\subset H_k$, and $G_S$ has constant singlet intrinsic torsion, then there is a consistent truncation of the supergravity theory to $Z$. 
\label{thm}
\end{thm}
\noindent
To unpack this a little, one first recalls that $M$ admits a generalised $G_S$-structure if the structure group of the bundle $\hat{E}$ can be taken to be a subgroup $G_S$ of the exceptional group. This is generally a topological obstruction. It allows one to decompose sections of $\hat{E}$ (and $\hat{N}$ and other bundles that carry representations of the exceptional group) into sub-bundles that transform in $G_S$ representations. In particular there may be sections that transform as singlets (that is are invariant). Suppose $G_S$ satisfies the conditions of the theorem and $\{K_\mathcal{A}{}^M(y)\}$ and $\{J_\Sigma{}^{I}(y)\}$ are bases, labelled by $\mathcal{A}$ and $\Sigma$, for the sets of $G_S$ singlets in $\hat{E}$ and $\hat{N}$ respectively. We can then make a truncation ansatz that keeps only these singlet degrees of freedom, that is, 
\begin{equation}
    \begin{aligned}
      \text{vectors:} &&& & \hat{\mathcal{A}}_\mu{}^M(x,y) &=
         A_\mu{}^{\mathcal{A}}(x) K_{\mathcal{A}}{}^M(y) , \\
      \text{two-forms:} &&& & \hat{\mathcal{B}}_{\mu\nu}{}^{I}(x,y) &=
         B_{\mu\nu}{}^\Sigma(x) J_\Sigma{}^{I}(y) , 
    \end{aligned}
    \label{eq:ct-gauge field}
\end{equation}
while the generalised metric can be written as 
\begin{equation}
    \hat{G}_{MN}(x,y) = \mathcal{V}_M{}^P(x)\mathcal{V}_N{}^Q(x) H_{PQ}(y) ,
    \label{eq:generalised-metric}
\end{equation}
where the $\mathcal{V}_M{}^P(x)$ are coset elements 
\begin{equation}
    \mathcal{V}_M{}^P(x) \in 
        \frac{C_{E_{k(k)}}(G_S)}{C_{H_k}(G_S)}:= \frac{\mathcal{G}}{\mathcal{H}} ,
    \label{eq:ct-scalar}
\end{equation}
embedded in $E_{k(k)}$, where $C_G(G')$ denotes the commutator (or centraliser) group of $G'\subset G$, and $H_{PQ}(y)$ is a fixed $\mathcal{H}\times G_S$ invariant generalised metric. On reducing to $Z$, we have scalar fields $\mathcal{V}_M{}^P(x)$, with sigma-model kinetic terms given by the metric on $\mathcal{G}/\mathcal{H}$, vector fields $\mathcal{A}_\mu{}^{\mathcal{A}}(x)$ and two-forms $\mathcal{B}_{\mu\nu}{}^\Sigma(x)$.

For a general $G_S$ structure the reduction is not consistent because the $G_S$-singlet degrees of freedom $\mathcal{V}_M{}^P$, $\mathcal{A}_\mu{}^{\mathcal{A}}$ and $\mathcal{B}_{\mu\nu}{}^\Sigma$ can  source non-singlet modes that we have not included in the truncation. These couplings arise from internal derivatives of the bases $\{K_\mathcal{A}{}^M(y)\}$ and $\{J_\Sigma{}^{I}(y)\}$ and $H_{PQ}(y)$. To ensure such couplings are not present we need to impose a differential condition on the $G_S$-structure. This is the ``singlet intrinsic torsion'' constraint mentioned in the theorem. It ensures that singlet modes can only couple to other singlet modes. The exact geometrical meaning of this condition is discussed in detail in~\cite{Cassani:2019vcl}. The structure of the reduced theory is then completely determined by the group theory of how $G_S$ embeds in $E_{k(k)}$ and which singlet intrinsic torsion terms are non-zero. In particular, the latter control the gauging of the theory and the scalar potential, so that if the whole intrinsic torsion vanishes the reduced theory is ungauged and the scalars are massless moduli. Furthermore, if the fermions also have singlet modes under $G_S$, then these should also be kept in the reduction and the reduced theory retains some degree of supersymmetry. 

\subsection{\texorpdfstring{$Spin(n,n)\times\mathbb{R}^+$ generalised geometry}{Spin(n,n) gg}}
\label{sec:Spin(n,n)-gg}

To see how the truncations of~\cite{Leung:2022nhy,Lin:2025ucv} fit into the generalised geometry framework we need to show that the brane solutions admit a generalised $G_S$ structure with singlet intrinsic torsion. We will find that this happens in a systematic way via an embedding of $Spin(n,n)\subset E_{n(n)}$ for eleven-dimensional supergravity and $Spin(n,n)\subset E_{n+1(n+1)}$ for type II supergravity, with the relevant $G_S$ structure being $Spin(n)\subset Spin(n,n)$. The exceptional generalised geometry reduces to a $Spin(n,n)$ version of generalised geometry first introduced in~\cite{Strickland-Constable:2013xta} and which we now summarise. (More details are given in Appendix~\ref{app:sgg}.) 

The generalised tangent space $E$ transforms in the chiral spinor representation of $Spin(n,n)$ and has the form 
\begin{equation}
    E\simeq TM\oplus\Lambda^{n-3}T^{*}M\oplus\left(\Lambda^{n}T^{*}M\otimes\Lambda^{n-5}T^{*}M\right)\oplus\left(\left(\Lambda^{n}T^{*}M\right)^{\otimes2}\otimes\Lambda^{n-7}T^{*}M\right)\oplus\cdots\,,
    \label{eq:generalised-bundle-E}
\end{equation}
We denote sections of $E$ by $V=v+\lambda+\dots$ or in components $V^M=(v^m,\lambda_{m_1\dots m_{m-3}},\dots)$. The corresponding adjoint bundle of $Spin(n,n)\times\mathbb{R}^+$ decomposes as
\begin{equation}
    \mathrm{ad}\simeq\mathbb{R}\oplus\left(TM\otimes T^{*}M\right)\oplus\Lambda^{n-2}TM\oplus\Lambda^{n-2}T^{*}M \, . 
    \label{eq:generalised-bundle-ad}
\end{equation}
Two other generalised tensor bundles will be of importance for us, namely 
\begin{align}
    P &\simeq\Lambda^{n}T^{*}M\oplus\Lambda^{2}T^{*}M\oplus\Lambda^{n-4}TM\oplus\left(\Lambda^{n}TM\otimes\Lambda^{n-6}TM\right)
    \oplus\cdots \, ,
    \label{eq:generalised-bundle-P} \\
    X &\simeq T^{*}M\oplus\Lambda^{n-1}T^{*}M\,.
    \label{eq:generalised-bundle-X}
\end{align}
where $P$ transforms in the spinor representation with the opposite chirality to $E$ and $X$ transforms in the $2n$-dimensional vector representation of $Spin(n,n)$. The $Spin(n,n)$ invariant inner product on $Y=\eta+\tilde{\eta}\in\Gamma(X)$ is given by 
\begin{equation}
    \langle Y, Y \rangle = \eta \wedge \tilde{\eta} \, , 
\end{equation}
where $\eta\in T^*M$ and $\tilde{\eta}\in\Lambda^{n-1}T^*M$. 

As always, the generalised tangent space parameterises the generalised Lie (or Dorfman) derivative which encodes  bosonic symmetries. In this case these are diffeomorphisms parameterised by $v\in\Gamma(TM)$ and gauge transformations of a $(n-2)$-form gauge potential $\tilde{A}$, parameterised by $\lambda\in\Gamma(\Lambda^{n-3}T^*M)$. Given $V\in\Gamma(E)$, the generalised Lie derivative act on generalised tensors as
\begin{equation}
    L_V = \mathcal{L}_v - \dd\lambda\; \cdot \, , 
    \label{eq:Dorfman-derivative-Spindd}
\end{equation}
where $\mathcal{L}_v$ is the ordinary Lie derivative and $\dd\lambda\in\Gamma(\mathrm{ad})$ acts via the $\mathfrak{so}(n,n)$ adjoint action. (All the other components of $E$ act trivially.) Changing frame by acting with the adjoint element $\ee^A$ on the generalised tensor bundles, the generalised Lie derivative takes the form, twisted by the field strength $\tilde{F}=\dd \tilde{A}$ and denoted $L^{\tilde{F}}_V$, 
\begin{equation}
    L^{\tilde{F}}_V = \mathcal{L}_v - (\dd\lambda-\imath_v \tilde{F}) \; \cdot \, . 
\end{equation}

Viewing the partial derivative as a section of $E^*$ defines a map, particular to $Spin(n,n)\times\mathbb{R}^+$ generalised geometry, 
\begin{equation}
    \dd : X \to P \, , 
    \label{eq:projection-P}
\end{equation}
such that $\eta + \tilde{\eta} \mapsto \dd\eta + \dd\tilde{\eta}$ where $\eta$ and $\tilde{\eta}$ are the one-form and $(n-1)$-form sections of $X$. The action is $Spin(n,n)$ covariant in that it commutes with the Dorfman derivative
\begin{equation}
    [L_V,\dd]=0 \, , 
\end{equation}
while written in the twisted frame the operation becomes 
\begin{equation}
    \dd_{\tilde{F}}(\eta+\tilde{\eta}) =  \dd\eta + \bigl( \dd\tilde{\eta} + \left(-1\right)^{\left[\frac{n-1}{2}\right]} \tilde{F}\wedge \eta \bigr) \,.
    \label{eq:dF}
\end{equation}

The maximally compact subgroup of $Spin(n,n)\times\mathbb{R}^+$ is $Spin(n)\times Spin(n)$, and as in other generalised geometries one can define a generalised metric $G\in \Gamma(S^2E^*)$ parametrising, at each point $y\in M$,the coset
\begin{equation}
    G|_y \ni \frac{Spin(n,n)}{Spin(n)\times Spin(n)}\times\mathbb{R}^+\,.
\end{equation}
Decomposing into conventional tensors, it defines a conventional metric $\tilde{g}_{mn}$, the $(n-2)$-form potential $\tilde{A}_{m_1m_2\cdots m_{n-2}}$ and a scalar $\tilde{\Delta}$. Concretely, given a generalised vector $V^M=(v^m,\lambda_{m_1\dots m_{d-3}},\dots)$, we have, in the untwisted basis, 
\begin{equation}
    \begin{aligned}
         G(V,V) &= G_{MN}V^M V^N  = \ee^{-(n-4)\tilde{\Delta}/2} \Bigl( \tilde{g}_{mn} v^m v^n \\
        &\quad+
        \tfrac{1}{(n-3)!}\tilde{g}^{m_1n_1}\cdots \tilde{g}^{m_{n-3}n_{n-3}}
        (\lambda - \imath_v \tilde{A})_{m_1\dots m_{n-3}}
        (\lambda - \imath_v \tilde{A})_{n_1\dots n_{n-3}} +\dots \Bigr) 
    \end{aligned}
    \label{eq:generalised metric}
\end{equation}
The $Spin(n)\times Spin(n)$ structure also defines a metric $G'$ on the $X$ bundle that given $Y=\eta+\tilde{\eta}\in\Gamma(X)$ takes the form 
\begin{equation}
\begin{aligned}
    G'(Y,Y) &= \ee^{-2\tilde{\Delta}} \Bigl( \tilde{g}^{mn} \eta_m \eta_n \\ & \qquad {}
        + \tfrac{1}{(n-1)!}\tilde{g}^{m_1n_1}\cdots \tilde{g}^{m_{n-1}n_{n-1}}
        (\tilde{\eta} - \eta \wedge s(\tilde{A}))_{m_1\dots m_{n-1}}
        (\tilde{\eta} - \eta \wedge s(\tilde{A}))_{n_1\dots n_{n-1}} \Bigr) \, , 
\end{aligned}
\label{eq:generalised metric-X}
\end{equation}
where $s(\alpha)=(-1)^{[p/2]}$ for $\alpha\in\Gamma(\Lambda^pT^*M)$. The different powers of $\ee^{\tilde{\Delta}}$ in the two metrics comes from the different $\mathbb{R}^+$ weights of the $E$ and $X$ bundles. One can view the $Spin(n)\times Spin(n)$ structure as defined by either metric\footnote{Note that this reparametrisation is degenerate for the $n=2$ case since for both $G$ and $G'$ the $\tilde{\Delta}$ dependence can be absorbed into a rescaled metric $g'_{mn}=\ee^{2\tilde{\Delta}}g_{mn}$ and so both appear to be completely determined by $\tilde{g}$ and the zero-form $\tilde{A}$. A non-degenerate parametrisation of $G'$ is given in~\eqref{eq:G'n=2}.\label{foot:n=2}} and in the following it will typically be more useful to use $G'$. In particular, given a conventional orthonormal frame $\{e^a\}$ for the metric $\tilde{g}$ we get an generalised ``split'' frame for $G'$ given by 
\begin{equation}
\label{eq:G'split-frame}
    E^A = (E^a, E'^a) \quad \text{where} \quad \left\{
    \begin{aligned}
        E^a &= \ee^{\tilde{\Delta}} \bigl( e^a + e^a \wedge s(\tilde{A}) \bigr) \\
        E'^a &= \ee^{\tilde{\Delta}}\, \tilde{*} e^a
    \end{aligned} \right.
\end{equation}
where $\tilde{*}$ is the Hodge dual defined by $\tilde{g}$, which satisfies the orthonormality conditions $G'(E^m,E^n)=G'(E'^m,E'^n)=\delta^{mn}$ and $G'(E^m,E'^n)=0$.

As in conventional geometry one can define a generalised connection $D$ as a differential operator acting on generalised vectors
\begin{equation}
    D_M V^N =\partial_M V^N + \Omega_M{}^N{}_P V^P \, , 
\end{equation}
where $\partial_M$ denotes the embedding of the partial derivative in $E^*$. We will always take $D$ to be compatible with the $Spin(n,n)\times\mathbb{R}^+$ structure in which case it extends to an action on any generalised tensor. As discussed in~\cite{Strickland-Constable:2013xta}, one can define the generalised torsion $T(D)$ of $D$ and this is a tensor transforming in the chiral spinor and chiral spin-$\frac32$ representations of $Spin(n,n)$. It is relatively easy to show that, given $Y\in\Gamma(X)$, the generalised torsion is encoded by  
\begin{equation}
\label{eq:gen-torsion}
    (\dd_D - \dd) Y = - T(D)\cdot Y \, ,
\end{equation}
where $\dd_DY$ is the projection of $DY\in\Gamma(E^*\otimes X)$ onto $\Gamma(P)$ and $T(D)\cdot Y$ is the tensor product of $T(D)$ and $Y$ projected onto $\Gamma(P)$.\footnote{This is the analogue of $2\nabla_{[m}\alpha_{n]}-2\partial_{[m}\alpha_{n]}=-T(\nabla)^p{}_{mn}\alpha_p$ in conventional geometry.} 
 
\subsection{Branes and torsion-free $Spin(n)\subset Spin(n,n)$ structures}

A generic flat brane solution, be it in type II or eleven-dimensional supergravity has a metric and a magnetic $(n-1)$-form flux of the form 
\begin{equation}
        \dd \hat{s}^2 = H^{2\alpha}(y)\eta_{\mu\nu}\dd x^\mu \dd x^\nu 
            + H^{2\beta}(y)\delta_{mn}\dd y^m \dd y^n\,, \qquad \hat{F} = -*_\delta\dd H(y) \, , 
    \label{eq:p-brane}
\end{equation}
where $x^\mu$ are coordinates on the brane worldvolume $Z$ and $y^m$ are coordinates on the transverse space $M\simeq \mathbb{R}^n$, and $*_\delta$ is the Hodge dual on the transverse space with the flat  metric $\delta_{mn}$. $H(y)$ is a harmonic function in the transverse space satisfying 
\begin{equation}
    \dd^* \dd H(y) = \partial_m\partial^m H(y) = 0\,.
\end{equation} 

Combining the metric $\tilde{g}_{mn}$ and $(n-2)$-form potential $\tilde{A}_{m-1\dots m_{n-2}}$ on the transverse space with an additional function $\tilde{\Delta}$ we see that we have the degrees of freedom of a generalised metric in $Spin(n,n)\times \mathbb{R}^+$ generalised geometry. The purpose of this subsection is to show that the transverse geometry of any brane solution~\eqref{eq:p-brane} defines a torsion-free generalised $Spin(n)\subset Spin(n,n)$ structure, which is the origin of the consistent truncation. 

Consider a $Spin(n,n)\times\mathbb{R}^+$ generalised metric with components $(\tilde{g},\tilde{A},\tilde{\Delta})$ such that\footnote{For the degenerate $n=2$ case the corresponding generalised metric $G'$ is still well-defined but we need to take a different parameterisation (see~\eqref{eq:G'n=2}).} 
\begin{equation}
    \dd \tilde{s}^2 = \tilde{g}_{mn}\dd y^m\dd y^n = H^{2/(n-2)} \delta_{mn}\dd y^m \dd y^n  \, , \qquad 
    \tilde{F} = - *_\delta\dd H \, , \qquad
    \ee^{2\tilde{\Delta}} = H^{-2/(n-2)} \, ,
    \label{eq:Spinnn-bg}
\end{equation}
where $H$ is harmonic. We see that the split frame~\eqref{eq:G'split-frame} for this the background is given by 
\begin{equation}
\label{eq:global-frame}
\begin{aligned}
    E^m &= \dd y^m + \dd y^m \wedge s(\tilde{A}) \\
    E'^m &=  H \, {*_\delta \dd y^m}   
\end{aligned}
\end{equation}
and is  globally defined. Thus the background actually admits an identity structure and is generalised parallelisable\footnote{This is a consequence of the fact that $M\simeq\mathbb{R}^n$ is parallelisable in the conventional sense.}. However, the identity structure does not have singlet torsion. Working in the untwisted picture (where the $s(\tilde{A})$ term in~\eqref{eq:global-frame} is absent and we use the $\dd_{\tilde{F}}$ operator) we have 
\begin{equation}
\begin{aligned}
    \dd_{\tilde{F}} E^m &= - (-1)^{[(n-1)/2]}\, {*_\delta\dd H}\wedge \dd y^m \, , \\
    \dd_{\tilde{F}} E'^m &= \dd H \wedge {*_\delta\dd y^m} \, ,
\end{aligned}
\end{equation}
and neither right-hand side is generically a sum of singlet generalised tensors with constant coefficients as would be required for singlet torsion. 

However the linear combinations 
\begin{equation}
    \Xi^m = \dd y^m + (-1)^{[n/2]}H \, {*_\delta \dd y^m}\,.
    \label{eq:Spin(n)-singlets}
\end{equation}
satisfy $\dd_{\tilde{F}}\Xi^m=0$. Since under the $SO(n,n)$ invariant norm we have $\langle \Xi^m,\Xi^n\rangle \propto \delta^{mn}$, these $n$ generalised vectors define a $G_S=Spin(n)_S\subset Spin(n)_S\times Spin(n)\subset Spin(n,n)$ structure group, where we use the subscript to indicate which factor gives the structure group $G_S$. (In the consistent truncation the second factor will form part of the R-symmetry group of the truncated theory.) In terms of representations 
\begin{equation}
    \mathbf{2n} = (\mathbf{n}, \mathbf{1}) \oplus (\mathbf{1},\mathbf{n})\, ,
    \label{eq:X-rep-decompose}
\end{equation}
and the $\Xi^m$ are the $n$ $G_S$-invariant objects in the second factor $(\mathbf{1},\mathbf{n})$. The spinor representations have no singlets under $G_S$ and hence there are no invariant generalised tensors in $E$. From~\eqref{eq:gen-torsion}, in the untwisted frame, taking $D$ to be a $G_S$-compatible connection implies $\dd_D\Xi^m=0$, and hence $\dd_{\tilde{F}}\Xi^m=0$ implies that the intrinsic torsion of $D$ vanishes, or in other words $\{\Xi^m\}$ define a torsion-free $Spin(n)$ structure and so can be the basis of a consistent truncation.\footnote{Since $P$ contains no $Spin(n)$ singlets, in fact the only way to have singlet intrinsic torsion in this case is, as here, for the whole intrinsic torsion to vanish.}

\section{Consistent truncation to supergravities on branes} \label{sec:ct on brane}

In this section, we derive the consistent truncations on the $p$-brane backgrounds following the general procedure of Sec.~\ref{sec:ct review}, by using the torsion-free $Spin(n)_S\subset Spin(n,n)$ structure defined by~\eqref{eq:Spin(n)-singlets}. In each case we need to understand how the $Spin(n,n)$ generalised geometry embeds into the $E_{k(k)}$ exceptional generalised geometry where $k=n+1$ for the type II supergravity solutions and $k=n$ for the eleven-dimensional M5-brane solution.  

For the scalar fields in the consistent truncation we note that 
\begin{equation}
    \frac{C_{Spin(n,n)}(Spin(n))}{C_{Spin(n)\times Spin(n)}(Spin(n))}\simeq \frac{Spin(n)}{Spin(n)}\simeq\mathbf{1}\,,
    \label{eq:trivial-commutant}
\end{equation}
and so any scalar degrees of freedom must arise from the embedding of the $Spin(n,n)$ generalised geometry into the $E_{k(k)}$. To see how these arise it is useful to identify at least a subset of fields in the embedding of the $Spin(n,n)$ degrees of freedom into the $E_{k(k)}$ degrees of freedom. From for example Appendix B of~\cite{Cassani:2016ncu}, we find that, in the Einstein frame, 
\begin{align}
    &\text{D$p$-brane ($p<7$):} &
    \dd s_E^2 &= \ee^{2\Delta_E} \dd s^2(Z) + \ee^{-\frac{(n-6)\Phi}{2(n-2)}} \dd \tilde{s}^2 \,, 
    \label{eq:Dp-brane-bg} \\
    &\text{NS-brane:} &
    \dd s_E^2 &= \ee^{2\Delta_E} \dd s^2(Z) + \ee^{-\Phi/2} \dd \tilde{s}^2 \,,
    \label{eq:NS-brane-bg} \\
    &\text{M5-brane:} &
    \dd s_M^2 &= \ee^{2\Delta_M} \dd s^2(Z) + \dd \tilde{s}^2 \, .
    \label{eq:M-brane-bg}    
\end{align}
where $\dd \tilde{s}^2$ is the metric defined by the $Spin(n,n)$ generalised metric as in~\eqref{eq:generalised metric} and~\eqref{eq:generalised metric-X}, and rather than keeping all the scalar degrees of freedom we have just focussed on $g_{mn}$ and the dilaton $\Phi$, where appropriate. Note that for the D5-brane expression (with $n=4$) has warp factor of $\ee^{\Phi/2}$ in front of $\dd s^2$, reflecting the fact that it and the NS5-brane expression are related by S-duality. The D7-brane case is special because of the degeneration of the parameterisation of $G'$ and we will discuss it separately below. The corresponding $E_{k(k)}$-invariant volume forms are given by
\begin{equation}
    \mathrm{vol}_{\hat{G}} = 
    \begin{cases}
        \sqrt{\tilde{g}}\,\ee^{-\frac{n(n-6)\Phi}{4(n-2)}} \ee^{(8-n)\Delta_E} \,, & \text{D$p$-brane}\,, \\
        \sqrt{\tilde{g}}\,\ee^{-\Phi} \ee^{4\Delta_E} \,,& \text{NS5-brane}\,, \\
        \sqrt{\tilde{g}}\,e^{4\Delta_M} \,,& \text{M5-brane}\,, 
    \end{cases}
    \label{eq:invariant volume}
\end{equation}
and are equal to the $Spin(n,n)$ invariant volume $\vol_G=\sqrt{\tilde{g}}\,\ee^{2\tilde{\Delta}}$. The dilaton in the embeddings~\eqref{eq:Dp-brane-bg} and~\eqref{eq:NS-brane-bg} should give an additional scalar in the consistent truncation. To see exactly how it should appear in the ansatz, we note that, as discussed in~\cite{Cassani:2019vcl}, the ansatz must be such that $\mathrm{vol}_{\hat{G}}$ is fixed. Thus we see that for the type II cases we should take 
\begin{equation}
\label{eeq:sigma}
    \Phi \rightarrow \Phi + \kappa \sigma \,,\qquad \Delta_E \rightarrow \tau \sigma\,,
\end{equation}
such that 
\begin{align}
    n(n-6)\kappa &= 4(n-2)(8-n)\tau\,, & &\text{D$p$-brane ($p<7$),} 
    \label{eq:Dpbrane-scalar} \\
    \kappa &= 4 \tau \,, & & \text{NS5-brane.}
    \label{eq:NSbrane-scalar}
\end{align}

For the vector and tensors in the truncations, we find that there are two different cases:
\begin{equation}
\begin{aligned}
    \text{vectors:} &&
    \hat{E} &\supset X \, , &
    \{ K_\mathcal{A} \} &\supset \{ \Xi^m\} & && &
    \text{ D$p$-branes, IIB NS-brane,} \\
    \text{tensors:} &&
    \hat{N} &\supset X \, , &
    \{ J_\Sigma \} &\supset \{ \Xi^m\} & && &
    \text{ IIA NS5-brane, M5-brane.} \\   
\end{aligned}
\end{equation}
In each case there may be other additional singlet vectors and tensors in the truncation transforming in trivial $Spin(n,n)$ representations, but it includes these as a minimum. To compare with the truncation ans\"atze of \cite{Leung:2022nhy,Lin:2025ucv} we need the corresponding field strengths. Given we are in the untwisted frame this are given by twisted operator $\dd_F$ defined in \eqref{eq:dF}, extended to an action on $Z\times M$ so that 
\begin{equation}
    \begin{aligned}
      \text{vectors:} &&& & \dd_{\tilde{F}}\mathcal{A}(x,y) &=
         \dd\mathcal{A}{}^{\mathcal{A}}(x)\wedge K_{\mathcal{A}}(y)\, , \\
      \text{two-forms:} &&& & \dd_{\tilde{F}}\mathcal{B}(x,y) &=
         \dd\mathcal{B}{}^\Sigma(x) \wedge J_\Sigma(y)\,. 
    \end{aligned}
    \label{eq:ct-field strength}
\end{equation}
where we have used the torsion-free conditions $\dd_F K_\mathcal{A} = \dd_F J_\Sigma = 0$ and the fact that on $Z$ the operator reduces to $\dd$, since the flux $F$ acts only on $M$. In the generalised geometry we are  in the democratic formalism~\cite{Bergshoeff:2001pv, Coimbra:2011nw, Coimbra:2011ky}, and so one need to impose the Einstein-frame duality relations on the type II and eleven-dimensional fluxes 
\begin{equation}
    \begin{aligned}
        \text{type II:} && &
        \begin{cases}
        \hat{F}_{n} = (-1)^{[n/2]} \ee^{(n-5)\Phi/2}\hat{*}\hat{F}_{10-n}\,,\\
        \hat{H}_7 = - \ee^{-\Phi}\hat{*}\hat{H}_{3}\,,     
        \end{cases} \\
        \text{M-theory:} && & \quad\hat{F}_7 = \hat{*}\hat{F}_4\,.
    \end{aligned}
    \label{eq:duality}
\end{equation}
in making the final comparison with~\cite{Leung:2022nhy,Lin:2025ucv}. 

In the remainder of this section, we construct the consistent truncation explicitly for each case in turn. We will continue to use a hat to denote ten- or eleven-dimensional fields, operators and $E_{n(n)}$ bundles, distinguishing them from the corresponding $Spin(n,n)$ objects.

\subsection{D3-brane}
For the D3-brane we have $n=6$ and the supergravity solution
\begin{equation}    
    \begin{aligned}
     \dd \hat{s}_E^2 &= H^{-\frac12} \eta_{\mu\nu}\dd x^\mu \dd x^\nu 
        + H^{\frac12} \delta_{mn}\dd y^m \dd y^n \,, \\        
    \hat{F}_5 &= -\, {*_\delta\dd H} + H^{-2} \dd H \wedge \mathrm{vol}_\eta \,, 
        \qquad \hat{\Phi} = 0 \,,
    \end{aligned}
    \label{eq:D3-bg}
\end{equation}
where $\mathrm{vol}_\eta=\dd x^0\wedge\dd x^1\wedge \dd x^2 \wedge\dd x^3$. The resulting consistent truncation is to pure ungauged four-dimensional $\mathcal{N}=4$ supergravity. 

The $Spin(6)_S$ structure embeds in $E_{7(7)}$ as 
\begin{equation}
    E_{7(7)}\supset SL(2,\mathbb{R})\times Spin(6,6) \supset SL(2,\mathbb{R})\times Spin(6)\times Spin(6)_S \,,
\end{equation}
and hence the scalars in the consistent truncation parameterise the coset
\begin{equation}
    \frac{C_{E_{7(7)}}(Spin(6)_S)}{C_{SU(8)}(Spin(6))_S}\simeq \frac{SL(2,\mathbb{R})\times Spin(6)}{SO(2)\times Spin(6)}\simeq\frac{SL(2,\mathbb{R})}{SO(2)}\,.
    \label{eq:moduli-D3}
\end{equation}
From~\eqref{eq:Dpbrane-scalar} we see that the dilaton $\hat{\Phi}$, without any deformation of the warp factor, is a scalar in the truncation. Hence, as one might expect, the $SL(2,\mathbb{R})/SO(2)$ scalar manifold is simply the axion-dilaton pair $(\hat{\Phi},\hat{A}_0)$, and the $SL(2,\mathbb{R})$ action is that of S-duality.  

For the vector embedding ansatz, we have the decomposition under $E_{7(7)}\supset SL(2,\mathbb{R})\times Spin(6,6)$
\begin{equation}
\arraycolsep=1.6pt
\def\arraystretch{1.2}
\begin{array}{ccccc}
    \hat{E} &\simeq & S \otimes (T^*\oplus\Lambda^5T^*) & \oplus & (T\oplus \Lambda^3T^*\oplus \Lambda^6T^*\otimes T^*) \\ 
    &\simeq & (S\otimes X) & \oplus & E \\
    \mathbf{56} &=& (\mathbf{2}, \mathbf{12}) & \oplus & (\mathbf{1}, \mathbf{32})
\end{array}
\end{equation}
where $S\simeq\mathbb{R}^2$ transforms as a doublet of $SL(2,\mathbb{R})$. We see that $X\subset \hat{E}$ and hence the singlet tensors $\Xi^m$ define a basis for vector fields in the consistent truncation. In fact these come in an $SL(2,\mathbb{R})$ doublet $\{K_\mathcal{A}\}$, with $\mathcal{A}=(a,m)$ and $a=1,2$. Writing $\mathcal{A}^{\mathcal{A}}_\mu=(A^m,A'^m)$, this gives
\begin{equation}
    \mathcal{A}_\mu^{\mathcal{A}}K_{\mathcal{A}} = \left(\begin{array}{c}A^m_\mu\\ A'^m_\mu\\\end{array}\right)\dd y_m - \left(\begin{array}{c}A^m_\mu\\ A'^m_\mu\\\end{array}\right) H*_\delta\dd y_m\,,
\end{equation}
with the corresponding field strengths 
\begin{equation}
    \dd_F (\mathcal{A}_\mu^{\mathcal{A}}K_{\mathcal{A}})= \left(\begin{array}{c}F^m_2\\F'^m_2\\\end{array}\right)\dd y_m - \left(\begin{array}{c}F^m_2\\ F'^m_2\\\end{array}\right) H*_\delta\dd y_m = \left(\begin{array}{c}\hat{H}_{2,1}\\\hat{F}_{2,1}\\\end{array}\right) + \left(\begin{array}{c}\hat{F}_{2,5}\\-\hat{H}_{2,5}\\\end{array}\right)\, .
    \label{eq:D3-vector-ansatz}
\end{equation}
The subscripts on $\hat{F}_{p,q}$ denote an external (world-volume) $p$-form wedged with an internal (transverse space) $q$-form and the relationship to the ten-dimensional fluxes is fixed by requiring the correct Bianchi identities for, for example $\hat{H}_{2,5}$  hold:
\begin{equation}
    \begin{aligned}
        &\dd \hat{H}_{2,5} = F'^m_2 \wedge \dd (H *_\delta\dd y_m) = \hat{F}^{(\text{in})}_5 \wedge F_2'^m \wedge  \dd y_m = \hat{F}^{(\text{in})}_5 \wedge \hat{F}_{2,1}
    \end{aligned}
\end{equation}
where $\hat{F}^{(\text{in})}_5$ is the purely transverse part of the background field strength \eqref{eq:D3-bg}. Using \eqref{eq:duality}, this indeed coincides with the equation of motion for $\hat{F}_3$. Finally we impose the duality conditions \eqref{eq:duality}. The duality relating $\hat{F}_{2,5}$ and $\hat{F}_{2,1}$ implies
\begin{equation}
    F'^m_2=-\ee^{-\phi}*_gF^m_2\,,
    \label{eq:D3-vector-related}
\end{equation}
where $*_g$ denotes the Hodge dual associated with the general four-dimensional metric $g_{\mu\nu}$ that appears in the consistent truncation. Hence, we  obtain six vectors labelled by $m$ transforming in the vector representation of $SO(6)$. We do not need to consider the $\hat{N}$ bundle, since two-form gauge fields in four dimensional spacetime are dual to scalars. 

Combining \eqref{eq:D3-vector-ansatz} and \eqref{eq:D3-vector-related}, we recover the embedding ansatz
\begin{equation}\begin{split}\label{general}
  &\dd \hat{s} ^2_E= H^{-\frac12} \,g_{\mu\nu}\dd x^\mu \dd x^\nu + H^{\frac12}\,\delta_{mn}\dd y^m \dd y^n\,,\qquad\hat{\Phi}=\phi\,,\qquad \hat{A}_0=\chi\,,\\
  &\hat{F}_{5}=-\mathrm{vol}_{g}\wedge\dd H^{-1}-\ast_\delta \dd H\,,\qquad
  \hat F_{3}=-\ee^{-\phi}\ast_g F_{2}^m\wedge \dd y^m\,,\qquad\hat H_{3}=F_{2}^m\wedge \dd y^m \,,
\end{split}\end{equation}
which agrees with the ansatz of \cite{Leung:2022nhy} up to a field rescaling. 

\subsection{D4-brane}
For the D4-brane we have $n=5$ and the supergravity solution
\begin{equation}
    \begin{aligned}
        &\dd \hat{s}^2_E=H^{-\frac38}\eta_{\mu\nu}\dd x^\mu \dd x^\nu + H^{\frac58}\delta_{mn}\dd y^m \dd y^n\,,\\
        &\hat{F}_4=-*_\delta\dd H\,,\qquad \ee^{\hat\Phi} = H^{-\frac14}\,.
    \end{aligned}
    \label{eq:D4-bg}
\end{equation}
The resulting consistent truncation is to pure ungauged five-dimensional $\mathcal{N}=4$ supergravity. 

The $Spin(5)_S$ structure embeds in $E_{6(6)}$ as 
\begin{equation}
    E_{6(6)}\supset Spin(5,5)\times\mathbb{R}^+ \supset Spin(5)_S\times Spin(5)\times\mathbb{R}^+\,.,
\end{equation}
and there is a single scalar in the consistent truncation parameterising the coset
\begin{equation}
    \frac{C_{E_{6(6)}}(Spin(5))}{C_{USp(8)}(Spin(5))}\simeq\frac{\mathbb{R}^+\times Spin(5)}{Spin(5)}\simeq\mathbb{R}^+\,.
\end{equation}
From \eqref{eq:Dpbrane-scalar}, we have the scalar ansatz
\begin{equation}
    \dd \hat{s}^2_E=H^{-\frac38}\ee^{\frac{5}{8\sqrt{6}} \sigma}g_{\mu\nu}\dd x^\mu \dd x^\nu + H^{\frac58}e^{-\frac{3}{8\sqrt{6}}\sigma}\delta_{mn}\dd y^m \dd y^n\,,\quad \ee^{\hat{\Phi}}=H^{-\frac14}\ee^{-\frac{9}{4\sqrt6} \sigma}\,,
    \label{eq:D4-scalar-ansatz}
\end{equation}
where the overall normalisation is chosen so that $\sigma$ has a canonical kinetic term in the truncated theory.

For the vector embedding ansatz, we have the decomposition under $E_{6(6)}\supset Spin(5,5)\times \mathbb{R}^+$
\begin{equation}
\arraycolsep=1.6pt
\def\arraystretch{1.2}
\begin{array}{rcccccc}
\hat{E}
& \simeq & \mathbb{R}
& \oplus & (T^* \oplus \Lambda^4 T^*)
& \oplus & (T \oplus \Lambda^2 T^* \oplus \Lambda^5 T^*) \\

& \simeq & \mathbb{R}
& \oplus & X
& \oplus & E \\

\mathbf{27}
& = & \mathbf{1}_{-4}
& \oplus & \mathbf{10}_2
& \oplus & \mathbf{16}_{-1}
\end{array}
\end{equation}
We see that $X\subset \hat{E}$ and hence the singlet tensors $\Xi^m$ define a basis vector fields in the consistent truncation. There is also an additional vector coming from the trivial bundle, which is simply the RR vector $\hat{A}_1$. Thus  $K_\mathcal{A}=(1,\Xi^m)$ and we have 
\begin{equation}
    \mathcal{A}_\mu^{\mathcal{A}}K_{\mathcal{A}} = A^0_\mu + A^m_\mu\dd y_m + A^m_\mu H*_\delta\dd y_m\, ,
\end{equation}
with corresponding field strengths  
\begin{equation}
    \dd_F (\mathcal{A}^{\mathcal{A}}K_{\mathcal{A}}) = F^0_2 + F^m_2 \wedge \dd y_m + F^m_2 \wedge H*_\delta\dd y_m = \hat{F}_{2,0} + \hat{H}_{2,1} - \hat{F}_{2,4}\,. 
    \label{eq:D4-form-ansatz1}
\end{equation}
The relation to the ten-dimensional fluxes is again fixed by requiring the correct Bianchi identities. In particular
\begin{equation}
    \dd \hat{F}_{2,4} = -F^m_2 \wedge \dd (H *_\delta\dd y_m) = \hat{F}_4 \wedge F_2^m \wedge  \dd y_m = \hat{F}_4 \wedge \hat{H}_{2,1}
\end{equation}
provides the equation of motion of $\hat{F}_{3,1}\,.$
Finally, the duality condition \eqref{eq:duality} relating $\hat{F}_{2,4}$ and $\hat{F}_{3,1}$ imposes
\begin{equation}
     \hat{F}_{3,1} = -e^{\frac{2}{\sqrt{6}}\sigma}*_g F_2^m \dd y_m\,,
     \label{eq:D4-form-related}
\end{equation}
where $*_g$ denotes the Hodge dual associated with the general five-dimensional metric $g_{\mu\nu}$ that appears in the consistent truncation. We do not need to consider the $\hat{N}$ bundle, since two-form gauge fields in five dimensional spacetime are dual to vectors. 

Combining \eqref{eq:D4-scalar-ansatz}, \eqref{eq:D4-form-ansatz1} and \eqref{eq:D4-form-related}, we recover
\begin{equation}
  \begin{aligned}
    \dd \hat{s}^{2}&=H^{-\frac38} \ee^{\frac{5}{8\sqrt{6}}\sigma}g_{\mu\nu}\left(x\right)\dd x^{\mu}\dd x^{\nu}+H^{\frac58}\ee^{-\frac{3}{8\sqrt{6}}\sigma}\delta_{mn}\dd y^{m}\dd y^{n}\,,\quad \ee^{\hat{\Phi}}=H^{-\frac14}\ee^{-\frac{9}{4\sqrt{6}}\sigma}\,,\\
    \hat{F}_{4}&=-*_{\delta}\dd H-e^{\frac{2}{\sqrt{6}}\sigma}*_{g}F_{2}^{m}\wedge \dd y^{m}\,,\quad\hat{H}_{3}=F_{2}^{m}\wedge\dd y^{m}\,,\quad\hat{F}_{2}=F^0_{2}\,,
  \end{aligned}
  \label{eq:5dN4-bosonic-ansatz}
\end{equation}
which agrees with the ansatz of \cite{Leung:2022nhy,Lin:2025ucv} up to a field rescaling.

\subsection{M5-brane}
For the M5-brane we have $n=5$ and the supergravity solution
\begin{equation}
    \begin{aligned}
        &\dd \hat{s}^2=H^{-\frac13}(y)\eta_{\mu\nu}\dd x^\mu \dd x^\nu + H^{\frac23}(y)\delta_{mn}\dd y^m \dd y^n\,,\\
        &\hat{F}_4=-*_\delta \dd H\,.
    \end{aligned}
    \label{eq:M5-bg}
\end{equation}
The resulting consistent truncation is to pure ungauged six-dimensional $\mathcal{N}=(2,0)$ supergravity.

The $Spin(5)_S$ structure embedding is straightforward since 
\begin{equation}
    E_{5(5)}\simeq Spin(5,5)\supset Spin(5)_S\times Spin(5)\,.
\end{equation}
There are no scalars in the consistent truncation since
\begin{equation}
    \frac{C_{E_{5(5)}}(Spin(5))}{C_{USp(4)\times USp(4)}(Spin(5))}\simeq\frac{Spin(5)}{Spin(5)}\simeq \mathbf{1}\,.
\end{equation}
Also, the generalised vector bundle $\hat{E}$ transforming in the spinor representation of $Spin(5,5)$ contains no $Spin(5)_S$ singlet. Therefore, no vectors are retained either.

For the two-form embedding ansatz, we have 
\begin{equation}
    \hat{N} \simeq  T^*M \oplus  \Lambda^4 T^*M \simeq X \, , 
\end{equation}
and hence the singlet tensors $\Xi^m$ define a basis for tensor fields in the consistent truncation. We have 
\begin{equation}
    \mathcal{B}^\Sigma_{\mu\nu}J_\Sigma=B^m_{\mu\nu}\dd y_m + B^m_{\mu\nu}H *_\delta\dd y^m\,,
\end{equation}
with corresponding field strengths
\begin{equation}
     \dd_F(\mathcal{B}^m_2J_m)=H^m_3\wedge \dd y_m + H^m_3\wedge H *_\delta\dd y_m = \hat{F}_{3,1} + \hat{F}_{3,4}\,.
    \label{eq:M5-form-ansatz}
\end{equation}
The duality condition \eqref{eq:duality} relates $\hat{F}_{3,1}$ to $\hat{F}_{3,4}$ and hence requires the six-dimensional two-forms to be anti-self-dual,
\begin{equation}
    H_3 = -*_g H_3\,.
    \label{eq:M5-duality-related}
\end{equation}

Combining \eqref{eq:M5-form-ansatz} and \eqref{eq:M5-duality-related}, we obtain
\begin{equation}
\begin{aligned}
    &\dd \hat s ^2= H^{-\frac13} g_{\mu\nu}\dd x^\mu \dd x^\nu + H^{\frac23}\delta_{mn}\dd y^m \dd y^n\,,\\
    &\hat F_{4}=\delta_{mn} H_{3}^m\wedge \dd y^n-\ast_\delta \dd H\,,\qquad H_{3}^m=-\ast_g H_{3}^m\,.
\end{aligned}
\label{eq:6d-M5-ansatz}
\end{equation}
which is exactly the ansatz given in \cite{Leung:2022nhy}.

\subsection{Type II fivebranes}

We now turn to the fivebranes in type IIA and IIB. We will focus primarily on NS5-brane, since they give different truncations in the two theories. At the end we discuss briefly the D5-brane truncation as the S-dual of the IIB NS5-brane truncation. 

For an NS5-brane we have $n=4$ and the supergravity solution
\begin{equation}
    \begin{aligned}
        &\dd \hat{s}^2=H^{-\frac14}\eta_{\mu\nu}\dd x^\mu \dd x^\nu + H^{\frac34}\delta_{mn}\dd y^m \dd y^n\,,\\
        &\hat{H}_3=-*_\delta \dd H\,,\qquad e^{\hat\Phi} = H^{\frac12}\,.
    \end{aligned}
    \label{eq:NS5-bg}
\end{equation}
The $Spin(4)_S$ structure can embed in $E_{5(5)}\simeq Spin(5,5)$ in inequivalent ways depending on how $Spin(4,4)$ embeds.\footnote{These lead to different $GL(4,\mathbb{R})\subset Spin(4,4)$ embeddings that give the type IIA and type IIB generalised tangent space decompositions into conventional tensors.} In particular, the spinor representation of $Spin(5,5)$ decomposes as 
\begin{equation}
\label{eq:Spin55-decomps}
    \mathbf{16} = \begin{cases}
        \mathbf{8}_s \oplus \mathbf{8}_c & \text{for $Spin(4,4)_{\text{IIA}}\subset Spin(5,5)$} \,, \\
        \mathbf{8}_v \oplus \mathbf{8}_s & \text{for $Spin(4,4)_{\text{IIB}}\subset Spin(5,5)$} \,, \\
    \end{cases}
\end{equation}
in the two cases, where $\mathbf{8}_v$, $\mathbf{8}_s$ and $\mathbf{8}_c$ are the vector, spinor and conjugate spinor representations of $Spin(4,4)$ respectively. The two decompositions are related by the triality symmetry of $Spin(4,4)$. In both cases structure then embeds as $Spin(4,4)\subset Spin(4)_S\times Spin(4)\subset Spin(4)_S$ such that, following~\eqref{eq:X-rep-decompose},
\begin{equation}
    \mathbf{8}_v = (\mathbf{4},\mathbf{1})\oplus (\mathbf{1},\mathbf{4}) \,.
\end{equation}
As we will see, the two different embeddings lead to different truncations, with, in particular, either tensor or vector fields.

\subsubsection*{IIA theory}

The IIA solution has a consistent truncation to six-dimensional $\mathcal{N}=(2,0)$ supergravity coupled to one additional anti-self-dual tensor multiplet as we now show.

The scalars are parameterised by the coset
\begin{equation}
    \frac{C_{E_{5(5)}}(Spin(4))}{C_{Spin(5)\times Spin(5)}(Spin(4))}\simeq\frac{Spin(1,5)}{Spin(5)}\,,
\end{equation}
where the structure embeds as $Spin(4)_S\subset Spin(4,4)_\text{IIA}\subset Spin(5,5)$. As we saw for the M5-brane, pure $\mathcal{N}=(2,0)$ has no scalar degrees of freedom. The five scalars in $Spin(1,5)/Spin(5)$ are the scalar components of the additional anti-self-dual tensor multiplet. Again they must arise entirely from the embedding of the $Spin(4,4)_\text{IIA}$ geometry into the $E_{5(5)}\simeq Spin(5,5)$ geometry. 

To identify the ansatz for the scalars we note first that one will be the $\sigma$ scalar identified in~\eqref{eq:NSbrane-scalar}. Decomposing the $E_{5(5)}$ adjoint bundle under $Spin(4,4)_\text{IIA}$ we have
\begin{equation}
\begin{aligned}
    \widehat{\text{ad}} &\simeq 
        \bigl(\mathbb{R}\oplus TM\otimes T^*M \oplus \Lambda^2T^*M \oplus \Lambda^2TM \bigr) \\ & \qquad \qquad {}
            \oplus (T^*M \oplus \Lambda^3T^*M) 
            \oplus (TM \oplus \Lambda^3TM) \\
        &\simeq \text{ad} \oplus X \oplus X^* \, , 
\end{aligned}    
\end{equation}
where $\text{ad}$ is the $Spin(4,4)_\text{IIA}$ adjoint bundle. The $T^*M$ and $\Lambda^3T^*M$ bundles in $X$ correspond to the action of the RR one- and three-forms $\hat{A}_1$ and $\hat{A}_3$. Similar to the gauge field, the singlets $\Xi^m$ define a ansatz for four additional scalar fields 
\begin{equation}
    \hat{F}_{1,1}+\hat{F}_{3,1} = \tfrac{1}{\sqrt{2}}\dd_F(\phi^m\Xi_m)
        = \tfrac{1}{\sqrt{2}}\dd\phi^m \wedge (\dd y_m + H\, {*_\delta \dd y_m} ) \,.
\end{equation}
Combining~\eqref{eq:NSbrane-scalar}, the scalar ansatz is 
\begin{equation}
    \begin{aligned}
        &\dd \hat{s}^2=H^{-\frac14}\ee^{-\frac{1}{2\sqrt{2}}\sigma}g_{\mu\nu}\dd x^\mu \dd x^\nu + H^{\frac34}(y)\ee^{\frac{1}{2\sqrt{2}}\sigma}\delta_{mn}\dd y^m \dd y^n\,,\qquad \ee^{\hat{\Phi}} = H^{\frac12}\ee^{-\frac{1}{\sqrt{2}}\sigma}\,,\\
        &\hat{F}_{1,1} =\tfrac{1}{\sqrt2} \dd\phi^m \wedge \dd y^m\,,\qquad \hat{F}_{3,1}= \tfrac{1}{\sqrt2} \dd\phi^m \wedge H{*_\delta \dd y_m} \,.
    \end{aligned}
    \label{eq:NS5A-scalar-ansatz}
\end{equation}
The overall normalisations of $\sigma$ and $\phi^m$ are chosen to match standard expressions for hyperbolic metric. (This is discussed in some detail in appendix~\ref{app:NS5A}.) The exponentiated action of $\hat{A}_1+\hat{A}_3$ is nilpotent and together $\sigma$ and $\phi^m$ define ``horospherical coordinates'' on the hyperbolic  space $Spin(5,1)/Spin(5)$ with $\sigma$ parameterising the real Cartan subalgebra and $\phi^m$ the nilpotent factor. 

As can be seen from~\eqref{eq:Spin55-decomps}, the generalised vector bundle $\hat{E}$ contains no $Spin(4)_S\subset Spin(4,4)_\text{IIA}$ singlets. Therefore there are no vectors in the consistent truncation. For the two-form embedding ansatz, we have the decomposition under $E_{5(5)}\supset Spin(4,4)\times \mathbb{R}^+$
\begin{equation}
\arraycolsep=1.6pt
\def\arraystretch{1.2}
\begin{array}{rcccccc}
\hat{N}
& \simeq & \mathbb{R}
& \oplus & (T^*M \oplus \Lambda^3 T^*M)
& \oplus & \Lambda^4 T^*M \\

& \simeq & \mathbb{R}
& \oplus & X
& \oplus & \Lambda^4 T^*M \\

\mathbf{10}
& = & \mathbf{1}_{4}
& \oplus & \mathbf{8}_{v,\mathbf{0}}
& \oplus & \mathbf{1}_{-2}
\end{array}
\end{equation}
and the corresponding basis of singlet 2-forms $\{J_\Sigma\}$ 
\begin{equation}
    J_0 = 1\,,\qquad J_m = \Xi_m=\dd y_m + H *_\delta \dd y_m\,,\qquad J_5=H\vol_\delta\,,
\end{equation}
where $\vol_\delta=\dd y^1\wedge\dd y^2\wedge\dd y^3\wedge\dd y^4$, all closed under $\dd_F$. Here, $H\vol_\delta$ is the $E_{k(k)}$ invariant given in \eqref{eq:invariant volume}. Thus
\begin{equation}
    \mathcal{B}^\Sigma_{\mu\nu}J_\Sigma=B^0_{\mu\nu} + B^m_{\mu\nu}\dd y_m + B^m_{\mu\nu}H *_\delta\dd y_m + B^5_{\mu\nu}H\vol_\delta\,,
\end{equation}
with the corresponding field strengths
\begin{equation}
\begin{aligned}
    \dd_F (\mathcal{B}^\Sigma_2J_\Sigma)
    &=H^0_3 + H^m_3\wedge\dd y_m + H^m_3 \wedge H\, {*_\delta\dd y_m} 
       + H^5_3\wedge H\vol_\delta \\
    &= \hat{H}_{3,0} + \hat{F}_{3,1} + \hat{F}_{3,3} + \hat{H}_{3,4}\,.
\end{aligned}
\label{eq:NS5A-form-ansatz}    
\end{equation}
The duality condition \eqref{eq:duality} relates $\hat{F}_{3,1}$ to $\hat{F}_{3,3}$ and $\hat{H}_{3,0}$ to $\hat{H}_{3,4}$ such that 
\begin{equation}
\label{eq:NS5A-bianchi-cond}
    H_3^m = *_g H_3^m \,, \qquad H_3^0 = -*_g H_3^5 \,. 
\end{equation}
where $H^0_3=\dd B^0_2$, $H^m_3=\dd B^m_2$ and $H^5_3=\dd B^5_2$. Hence, in the consistent truncation, one obtains four self-dual two-forms $B^m_{\mu\nu}$ in the vector representation of $SO(4)$ together with one singlet two-form $B^0_{\mu\nu}$. The former, together with the self-dual part of $B^0_{\mu\nu}$, encoded in the supergravity multiplet, while the anti-self-dual part of $B^0_{\mu\nu}$ forms the extra tensor multiplet.

Combining~\eqref{eq:NS5A-scalar-ansatz}, \eqref{eq:NS5A-form-ansatz} and~\eqref{eq:NS5A-bianchi-cond} we obtain
\begin{equation}
    \begin{aligned}
        \dd\hat{s}^{2}=&H^{-\frac14}\ee^{-\frac{1}{2\sqrt{2}}\sigma}g_{\mu\nu}\dd x^{\mu}\dd x^{\nu}+H^{\frac34}\ee^{\frac{1}{2\sqrt{2}}\sigma}\delta_{mn}\dd y^{m}\dd y^{n}\,,\\
        \hat{H}_{3}=&-*_{\delta}\dd H+H_{3}^{0}\,,\quad
        \hat{F}_{2}=\tfrac{1}{\sqrt{2}}\dd\phi^{m}\wedge\dd y_{m}\,,\quad 
        \ee^{\hat{\Phi}}=H^{\frac{1}{2}}e^{-\frac{1}{\sqrt{2}}\sigma}\,,\\
        \hat{F}_{4}=&H_{3}^{m}\wedge\dd y^{m}
            +\tfrac{1}{\sqrt{2}} H\dd\phi^{m} \wedge {*_\delta\dd y_m}\,,
    \end{aligned}
    \label{eq:NS5A-embedding-ansatz}
\end{equation}
where $H^0_3=\dd B^0_2\,,\ H^m_3=\dd B^m_2$, and $H^m_3=*_{g}H^m_3$. Making a further truncation to pure $\mathcal{N}=(2,0)$ supergravity we see this agrees with the ansatz of~\cite{Leung:2022nhy}. A direct proof of consistency, rather than via generalised geometry, of the theory coupled to an additional tensor multiplet has not been given before and for completeness we include it here in App.~\ref{app:NS5A}.

\subsubsection*{IIB theory}
The IIB solution has a consistent truncation to pure six-dimensional $\mathcal{N}=(1,1)$ supergravity as we now show. Under $E_{5(5)}\supset SO(4,4)_{\text{IIB}}\supset SO(4)_S$, there is only one scalar in the truncation
\begin{equation}
    \frac{C_{E_{5(5)}}(Spin(4))}{C_{USp(4)\times USp(4)}(Spin(4))}\simeq\frac{\mathbb{R}^+\times Spin(4)}{Spin(4)}\simeq \mathbb{R}^+\,.
\end{equation}
From \eqref{eq:NSbrane-scalar}, we have the scalar ansatz
\begin{equation}
    \dd \hat{s}^2=H^{-\frac14}\ee^{-\frac{1}{2\sqrt{2}}\sigma}g_{\mu\nu}\dd x^\mu \dd x^\nu + H^{\frac34}\ee^{\frac{1}{2\sqrt{2}}\sigma}\delta_{mn}\dd y^m \dd y^n\,,\qquad \ee^{\hat{\Phi}} = H^{\frac12}\ee^{-\frac{1}{\sqrt{2}}\sigma}\,.
    \label{eq:NS5B-scalar-ansatz}
\end{equation}
where the overall normalisation is chosen so that $\sigma$ has a canonical kinetic term in the truncated theory.

For the vector embedding ansatz, we have the decomposition under $E_{5(5)}\supset Spin(4,4)\times \mathbb{R}^+$
\begin{equation}
\arraycolsep=1.6pt
\def\arraystretch{1.2}
\begin{array}{rcccccc}
\hat{E}
& \simeq & (TM \oplus T^*M)
& \oplus & (T^*M \oplus \Lambda^3 T^*M) \\

& \simeq & E
& \oplus & X \\

\mathbf{16}
& = & \mathbf{8}_{s,\mathbf{-1}}
& \oplus & \mathbf{8}_{v,\mathbf{1}}
\end{array}
\label{eq:IIB-E-bundle}
\end{equation}
and hence the singlet tensors $\Xi^m$ define a basis vector fields in the consistent truncation. We have
\begin{equation}
    \mathcal{A}^\mathcal{A}K_\mathcal{A}=A^m_{\mu}\dd y_m + A^m_{\mu}H *_\delta\dd y_m \,.
\end{equation}
with the corresponding field strengths
\begin{equation}
    \dd_F (\mathcal{A}^\mathcal{A}K_\mathcal{A}) = F^m_2\dd y_m + F^m_2 H *_\delta\dd y_m = \hat{F}_{2,1} + \hat{F}_{2,3}\,.
    \label{eq:NS5B-vector-ansatz}
\end{equation}
The duality condition~\eqref{eq:duality} imposes no further relations and hence we have four vectors in the vector representation of $SO(4)$, forming part of the supergravity multiplet.

For the two-form embedding ansatz, we have the decomposition under $E_{5(5)}\supset Spin(4,4)_{\text{IIB}}\times \mathbb{R}^+$
\begin{equation}
\arraycolsep=1.6pt
\def\arraystretch{1.2}
\begin{array}{rcccccc}
\hat{N}
& \simeq & \mathbb{R}
& \oplus & (\mathbb{R}\oplus\Lambda^2 T^*M\oplus \Lambda^4T^*M)
& \oplus & \Lambda^4T^*M\\

& \simeq & \mathbb{R}
& \oplus & P
& \oplus & \Lambda^4T^*M\\

\mathbf{10}
& = & \mathbf{1}_{2}
& \oplus & \mathbf{8}_{c,\mathbf{0}}
& \oplus & \mathbf{1}_{-2}
\end{array}
\label{eq:IIB-N-bundle}
\end{equation}
One can choose the singlet two-form basis to be
\begin{equation}
    J_0 = 1\,,\qquad J_1=H\vol_\delta\,,
\end{equation}
where $H\vol_\delta$ is the invariant volume \eqref{eq:invariant volume}, so that 
\begin{equation}
    \mathcal{B}^\Sigma_{\mu\nu}J_\Sigma=B^0_{\mu\nu} + B^1_{\mu\nu}H\vol_\delta \,,
    \label{eq:NS5B-form-ansatz}
\end{equation}
with the corresponding field strengths
\begin{equation}
    \dd (\mathcal{B}^\Sigma_2 J_\Sigma)
        = H^0_3 + H^1_3\wedge H\vol_\delta = \hat{H}_{3,0} + \hat{H}_{3,4} \, . 
\end{equation}
The duality condition \eqref{eq:duality} relates 
\begin{equation}
\label{eq:NS5B-bianchi-cond}
    H_3^0 = -*_g H_3^1 \,,
\end{equation}
leaving one independent two-form $B^0_{\mu\nu}$ in the trivial representation of $SO(4)$. It belongs to the supergravity multiplet.

Combining \eqref{eq:NS5B-scalar-ansatz}, \eqref{eq:NS5B-vector-ansatz} and \eqref{eq:NS5B-form-ansatz}, we obtain
\begin{equation}
  \begin{aligned}
    \dd \hat{s}^{2}=&H^{-\frac14}\ee^{-\frac{\sqrt{2}}{4}\sigma}g_{\mu\nu}\dd x^{\mu}\dd x^{\nu}+H^{\frac34}\ee^{\frac{\sqrt{2}}{4}\sigma}\delta_{mn}\dd y^{m}\dd y^{n}\,,\\
    \hat{H}_{3}=&-*_{\delta}\dd H+ H^0_{3}\,,\quad
    \hat{F}_{3}=F_{2}^{m}\wedge\dd y_{m}\,,\quad 
    \ee^{\hat{\Phi}}=H^{\frac12}\ee^{-\frac{1}{\sqrt{2}}\sigma}\,,\\
    \hat{F}_{5}=& F_{2}^{m}\wedge H{*_\delta}\dd y_m - \ee^{-\frac{1}{\sqrt{2}}\sigma}*_{g}F_{2}^{m}\wedge\dd y^{m}\,.
  \end{aligned}
  \label{eq:NS5B-embedding-ansatz}
\end{equation}
where the last term in $\hat{F}_5$ is fixed by self-duality. This agrees with the ansatz of \cite{Lin:2025ucv} up to a rescaling.

Finally let us mention the truncation on the D5-brane. The background is S-dual to the NS5-brane background, so takes the form~\eqref{eq:NS-brane-bg} with $\hat{F}_3$ replacing $\hat{H}_3$ and $\ee^{\hat{\Phi}}=H^{-1/2}$. As discussed in~\cite{Lin:2025ucv}, making an S-duality transformation on the truncation ansatz~\eqref{eq:NS5B-embedding-ansatz} gives the D5-brane ansatz
\begin{equation}
  \begin{aligned}
    \dd \hat{s}^{2}=&H^{-\frac14}\ee^{-\frac{\sqrt{2}}{4}\sigma}g_{\mu\nu}\dd x^{\mu}\dd x^{\nu}+H^{\frac34}\ee^{\frac{\sqrt{2}}{4}\sigma}\delta_{mn}\dd y^{m}\dd y^{n}\,,\\
    \hat{H}_{3}=& - F_{2}^{m}\wedge\dd y_{m}\,,\quad 
    \hat{F}_{3}= -*_{\delta}\dd H+ H^0_{3}\,,\quad
    \ee^{\hat{\Phi}}=H^{-\frac12}\ee^{\frac{1}{\sqrt{2}}\sigma}\,,\\
    \hat{F}_{5}=& F_{2}^{m}\wedge H{*_\delta}\dd y_m - \ee^{-\frac{1}{\sqrt{2}}\sigma}*_{g}F_{2}^{m}\wedge\dd y^{m}\,.
  \end{aligned}
  \label{eq:D5-embedding-ansatz}
\end{equation}

\subsection{D6-brane}
For a D6-brane we have $n=3$ and the supergravity solution
\begin{equation}
    \begin{aligned}
        &\dd \hat{s}^2=H^{-\frac18}\eta_{\mu\nu}\dd x^\mu \dd x^\nu + H^{\frac78}\delta_{mn}\dd y^m \dd y^n\,,\\
        &\hat{F}_2=-*_\delta \dd H\,,\qquad e^{\hat{\Phi}}(y) = H(y)^{-\frac34}\,.
    \end{aligned}
    \label{eq:D6-bg}
\end{equation}
The resulting consistent truncation is to pure ungauged seven-dimensional half-maximal supergravity. 

The $Spin(3)_S$ structure embeds in $E_{4(4)}$ as 
\begin{equation}
    E_{4(4)}\simeq SL(5,\mathbb{R})\supset SL(4,\mathbb{R})\times\mathbb{R}^+\simeq Spin(3,3)\times\mathbb{R}^+\supset Spin(3)_S\times Spin(3)\,.
\end{equation}
and there is a single scalar in the consistent truncation parameterising the coset
\begin{equation}
    \frac{C_{E_{4(4)}}(Spin(3))}{C_{Spin(5)}(Spin(3))}\simeq\frac{\mathbb{R}^+\times Spin(3)}{Spin(3)}\simeq\mathbb{R}^+\,.
\end{equation}
From \eqref{eq:Dpbrane-scalar} we can construct the embedding ansatz 
\begin{equation}
    \dd \hat{s}^2=H^{-\frac18}\ee^{-\frac{9}{8\sqrt{10}}\sigma}g_{\mu\nu}\dd x^\mu \dd x^\nu + H^{\frac78}\ee^{\frac{15}{8\sqrt{10}}\sigma}\delta_{mn}\dd y^m \dd y^n\,,\qquad \ee^{\hat{\Phi}} = H^{-\frac34}e^{\frac{5\sigma}{4\sqrt{10}}}\,.
    \label{eq:D6-scalar-ansatz}
\end{equation}
where the overall normalisation is chosen so that $\sigma$ has a canonical kinetic term in the truncated theory.

To determine the vector embedding ansatz, we have the decomposition under $E_{4(4)}\supset Spin(3,3)\times \mathbb{R}^+$
\begin{equation}
\arraycolsep=1.6pt
\def\arraystretch{1.2}
\begin{array}{rcccccc}
\hat{E}
& \simeq & (TM\oplus \mathbb{R} ) 
& \oplus & ( T^*M \oplus \Lambda^2T^*M ) \\

& \simeq & E
& \oplus & X \\

\mathbf{10}
& = & \mathbf{4}_{-3} & \oplus & \mathbf{6}_{2}

\end{array}
\end{equation}
and hence the singlet tensors $\Xi^m$ define a basis vector fields in the consistent truncation. We have
\begin{equation}
    \mathcal{A}^\mathcal{A}_{\mu}K_{\mathcal{A}}=A^m_{\mu}\dd y_m - A^m_{\mu}H *_\delta\dd y_m\,.
\end{equation}
with the corresponding field strengths
\begin{equation}
    \dd_F \left(\mathcal{A}^\mathcal{A}_{\mu}K_{\mathcal{A}}\right)=F^m_2\dd y_m - F^m_2H *_\delta\dd y_m = \hat{H}_{2,1} + \hat{F}_{2,2}\,.
    \label{eq:D6-vector-ansatz}
\end{equation}
The duality condition~\eqref{eq:duality} imposes no further relations and hence we have three vectors in the vector representation of $SO(3)$, forming part of the supergravity multiplet.

For the two-form embedding ansatz, we have the decomposition of the $\hat{N}$ bundle
\begin{equation}
\arraycolsep=1.6pt
\def\arraystretch{1.2}
\begin{array}{rcccccc}
\hat{N}
& \simeq & \Lambda^3T^*M
& \oplus & (\mathbb{R} \oplus T^*M) \\

& \simeq & \Lambda^3T^*M
& \oplus & E^* \\

\mathbf{5}
& = & \mathbf{1}_{-4}
& \oplus & \mathbf{4}_{1}
\end{array}
\end{equation}
and so we have a two-form singlet
\begin{equation}
    J = H\vol_\delta\,,
\end{equation}
which is the invariant volume \eqref{eq:invariant volume}, giving 
\begin{equation}
    \mathcal{B}^\Sigma_{\mu\nu}J_\Sigma = B_{\mu\nu}H\vol_\delta\,,
    \label{eq:D6-form-ansatz}
\end{equation}
and corresponding field strengths
\begin{equation}
   \dd (\mathcal{B}^\Sigma J_\Sigma)
      = H_3\wedge H\vol_\delta = \hat{F}_{3,3}
\end{equation}
The duality condition~\eqref{eq:duality} imposes no further relations (simply relating $\hat{F}_{3,3}$ to $\hat{F}_{4,0}$) and hence we have a single tensor field in the trivial representation of $SO(3)$, belonging to the supergravity multiplet.

Combining \eqref{eq:D6-scalar-ansatz}, \eqref{eq:D6-vector-ansatz}, and \eqref{eq:D6-form-ansatz}, we obtain the consistent truncation 
\begin{equation}
    \begin{aligned}
        &\dd \hat{s}^2=H^{-\frac18}\ee^{-\frac{9}{8\sqrt{10}}\sigma}g_{\mu\nu}\dd x^\mu \dd x^\nu + H^{\frac78}\ee^{\frac{15}{8\sqrt{10}}\sigma}\delta_{mn}\dd y^m \dd y^n\,,\qquad \ee^{\hat{\Phi}} = H^{-\frac34}e^{\frac{5\sigma}{4\sqrt{10}}}\,,\\
        &\hat{F}_2= -*_\delta \dd H\,,\quad \hat{H}_3=F^m_2\wedge \dd y_m\,,\quad \hat{F}_4=-F^m_2 H*_\delta \dd y_m +  e^{-\frac{4\sigma}{\sqrt{10}}}*_g  H_3 \,.
    \label{eq:D6-ansatz}
    \end{aligned}
\end{equation}
which is a new result. 

\subsection{D7-brane}
For the D7-brane we have $n=2$ and the supergravity solution 
\begin{equation}
    \begin{aligned}
        &\dd \hat{s}^{2}=\eta_{\mu\nu}dx^{\mu}dx^{\nu}+H\delta_{mn}\dd y^m\dd y^n  \,, \\
        &\hat{F}_1 = -*_\delta \dd H\,, \qquad \ee^{\hat{\Phi}}=H^{-1} \,,
    \end{aligned}
    \label{eq:D7-bg}
\end{equation}
and the truncation should be to pure $\mathcal{N}=1$ eight-dimensional supergravity. 

The $Spin(2)_S$ structure embeds in $E_{3(3)}\simeq SL(3,\mathbb{R})\times SL(2,\mathbb{R})$ as 
\begin{equation}
\begin{aligned}
    E_{3(3)} &\simeq SL(3,\mathbb{R})\times SL(2,\mathbb{R}) \\
    &\supset Spin(2,2) \simeq SL(2,\mathbb{R})\times SL(2,\mathbb{R}) \\
    &\supset Spin(2)_{S}\times Spin(2)\,.    
\end{aligned}
\label{eq:E3-decompose}
\end{equation}
where $Spin(2)_S$ embeds diagonally in $SL(2,\mathbb{R})\times SL(2,\mathbb{R})$. There is a single scalar in the consistent truncation parameterising the coset
\begin{equation}
\label{eq:D7-scalar}
    \frac{C_{SL(3,\mathbb{R})\times SL(2,\mathbb{R})}(Spin(2))}{C_{Spin(3)\times Spin(2)}(Spin(2))}\simeq\mathbb{R}^+\,.
\end{equation}

As mentioned in footnote~\ref{foot:n=2}, the parameterisation of the $Spin(n,n)$ generalised metric in~\eqref{eq:generalised metric} and~\eqref{eq:generalised metric-X} is degenerate for $n=2$. We can instead use the parameterisation 
\begin{equation}
    G'(Y,Y) = \tilde{h}^{mn} \bigl( \ee^{\tilde{\Sigma}} \eta_m\eta_n 
        + \ee^{-\tilde{\Sigma}} (\tilde{\eta} - \tilde{A}\eta)_m(\tilde{\eta} - \tilde{A}\eta)_n \bigr) \,. 
\label{eq:G'n=2}
\end{equation}
where the D7-brane background~\eqref{eq:D7-bg} is given by\footnote{This identification comes from comparing with Appendix E of~\cite{Ashmore:2015joa}. Note that unlike all the previous cases the dilaton is completely contained in $Spin(n,n)$ generalised metric.}
\begin{equation}
    \tilde{h}_{mn} = H\delta_{mn} \,, \qquad 
    \ee^{\tilde{\Sigma}} = \ee^{\hat{\Phi}} = H^{-1} \,, \qquad
    \tilde{A} = -\ee^{\tilde{\Sigma}} \hat{A}_0 \,. 
\end{equation}

To identify the scalar ansatz, we note that the $Spin(2,2)$-invariant volume form is $\vol_{G}=\sqrt{\tilde{h}}$. It is equal to the $E_{3(3)}$-invariant form 
\begin{equation}
    \vol_{\hat{G}} = \sqrt{g_E}\,\ee^{6\Delta_E} \,. 
\end{equation}
Since $\tilde{h}$ contain no scalar degrees of freedom and the invariant volume form cannot change, we see that the scalar $\sigma$ corresponds to 
\begin{equation}
    g_E \to \ee^{2\kappa\sigma} g_E\,,  \quad \Delta_E \to \Delta_E + \tau\sigma 
\end{equation}
with $\kappa=-3\tau$. Thus the scalar ansatz is 
\begin{equation}
   \dd \hat{s}^2=\ee^{-\frac{\sigma}{2\sqrt{3}}}g_{\mu\nu}\dd x^\mu \dd x^\nu + H\ee^{\frac{\sqrt{3}\sigma}{2}}\delta_{mn}\dd y^m \dd y^n\,,\qquad e^{\hat{\Phi}} = H^{-1}\,.
    \label{eq:D7-scalar-ansatz}
\end{equation}
where we have fixed the overall factor by requiring a standard kinetic term.

To determine the vector embedding ansatz, we have the decomposition under $E_{3(3)}\simeq SL(3,\mathbb{R})\times SL(2,\mathbb{R})\supset Spin(2,2)\simeq SL(2,\mathbb{R})\times SL(2,\mathbb{R})$
\begin{equation}
\arraycolsep=1.6pt
\def\arraystretch{1.2}
\begin{array}{rcccccc}
\hat{E}
& \simeq & TM 
& \oplus & ( T^*M \oplus T^*M ) \\
& \simeq & E
& \oplus & X \\
(\mathbf{3},\mathbf{2})
& = & (\mathbf{1},\mathbf{2}) & \oplus & (\mathbf{2},\mathbf{2})
\end{array}
\end{equation}
and hence the singlet tensors $\Xi^m$ define a basis vector fields in the consistent truncation. We have
\begin{equation}
    \mathcal{A}^\mathcal{A}_{\mu}K_{\mathcal{A}}=A^m_{\mu}\dd y_m - A^m_{\mu}H *_\delta\dd y_m\,.
\end{equation}
with the corresponding field strengths
\begin{equation}
    \dd_F \left(\mathcal{A}^\mathcal{A}_{\mu}K_{\mathcal{A}}\right)=F^m_2\dd y_m - F^m_2H *_\delta\dd y_m = \hat{H}_{2,1} + \hat{F}_{2,1}\,.
    \label{eq:D7-vector-ansatz}
\end{equation}
The duality condition~\eqref{eq:duality} imposes no further relations and hence we have two vectors in the vector representation of $SO(2)$, forming part of the supergravity multiplet.

For the two-form embedding ansatz, we have the decomposition of the $\hat{N}$ bundle
\begin{equation}
\arraycolsep=1.6pt
\def\arraystretch{1.2}
\begin{array}{rcccccc}
\hat{N}
& \simeq & (\mathbb{R}\oplus \mathbb{R}) & \oplus & \Lambda^2T^*M \\
(\mathbf{3},\mathbf{1})
& = & (\mathbf{2},\mathbf{1})
& \oplus & (\mathbf{1},\mathbf{1}) 
\end{array}
\end{equation}
The first term is a doublet of the $SO(2)_S$ structure (since it embeds diagonally in $SL(2,\mathbb{R})\times SL(2,\mathbb{R})$) and so we have a single two-form singlet
\begin{equation}
    J = H \vol_\delta\,.
\end{equation}
giving 
\begin{equation}
    \mathcal{B}^\Sigma_{\mu\nu}J_\Sigma = B_{\mu\nu} H\vol_\delta\,,
    \label{eq:D7-form-ansatz}
\end{equation}
and corresponding field strengths
\begin{equation}
   \dd (\mathcal{B}^\Sigma J_\Sigma)
      = H_3\wedge H\vol_\delta = \hat{F}_{3,2}
\end{equation}
The duality condition~\eqref{eq:duality} on imposes no further relations and hence we have a single tensor field in the trivial representation of $SO(2)$, belonging to the supergravity multiplet.

Combining \eqref{eq:D7-scalar-ansatz}, \eqref{eq:D7-vector-ansatz}, and \eqref{eq:D7-form-ansatz}, we obtain the consistent truncation 
\begin{equation}
    \begin{aligned}
        &\dd \hat{s}^2=\ee^{-\frac{\sigma}{2\sqrt{3}}}g_{\mu\nu}\dd x^\mu \dd x^\nu + H\ee^{\frac{\sqrt{3}\sigma}{2}}\delta_{mn}\dd y^m \dd y^n\,,\qquad e^{\hat{\Phi}} = H^{-1}\,,\\
        &\hat{F}_1= -*_\delta \dd H\,,\quad 
        \hat{H}_3=F^m_2\wedge \dd y_m\,,\quad 
        \hat{F}_3=-F^m_2 \wedge H*_\delta \dd y_m\,, \\
        &\hat{F}_5= H_3 \wedge H \vol_\delta + e^{-\frac{2\sigma}{\sqrt{3}}} *_g H_3 \,.
    \label{eq:D7-ansatz}
    \end{aligned}
\end{equation}
which is a new result.

\section{Conclusion and outlook}\label{sec:conclusion}

In this note we have shown how the consistent truncations of~\cite{Leung:2022nhy,Lin:2025ucv} based on supersymmetric brane solutions fit in the general analysis of~\cite{Cassani:2019vcl}. In particular, each brane solution defines a natural torsion-free generalised $Spin(n)$ structure in the $Spin(n,n)$ generalised geometry of~\cite{Strickland-Constable:2013xta}. Embedding this structure into $E_{n+1(n+1)}$ and $E_{n(n)}$ for the type II and eleven-dimensional supergravity respectively then defines the consistent truncation. We derived the truncations for D$p$-branes with $3\leq p\leq 7$, the M5-brane and the IIA and IIB NS5-branes. For the IIA NS5-brane and the D6- and D7-brane these represent new results. For completeness, in Appendix~\ref{app:NS5A}, we checked the consistency of the IIA NS5-brane ansatz directly without invoking exceptional generalised geometry. It is worth noting that although we ignored the fermion degrees of freedom here, their truncation ansatz again follows directly from the $Spin(n)$ structure and exceptional generalised geometry implies consistency~\cite{Cassani:2019vcl}. Although we have not done it here, one could then also match the fermion ans\"atze of~\cite{Lin:2025ucv}. 

In all cases other than the IIA NS5-brane the reduced theory was that of the appropriate pure half-maximal supergravity, while for the IIA NS5-brane one gets six-dimesional $\mathcal{N}=(2,0)$ supergravity coupled to a single tensor multiplet. Given the IIA NS5-brane is simply the reduction of the M5-brane solution and the consistent truncation on the latter gave pure $\mathcal{N}=(2,0)$ supergravity one might wonder where the extra tensor multiplet comes from. 

This again can be understood from the exceptional generalised geometry and indicates that there is a simple further extension of the consistent truncations we described here. For the M5-brane the transverse space is five-dimensional so that $n=5$ and since $E_{5(5)}\simeq Spin(5,5)$ the $Spin(n,n)$ generalised geometry is actually the same as the exceptional generalised geometry. As we have seen, for a generic harmonic function $H$, the generalised tensors given in~\eqref{eq:Spin(n)-singlets}, namely
\begin{equation}
    \Xi^m = \dd y^m + H\, {*_\delta \dd y^m} \in \Gs{X} \, , 
\end{equation}
with $n=1,\dots,5$ are both globally defined and torsion free, in that $\dd_F\Xi^m=0$. The IIA NS5-brane is a ``vertical reduction''~\cite{Lu:1996mg} of the M5-brane, that is one takes the M5-brane to be ``smeared'' so that $H$ is independent of, say, $y^5$ which then becomes the M-theory circle. To match the IIA NS5-brane solution, after this smearing we must have a torsion free $Spin(4)\subset E_{4(4)}\simeq Spin(5,5)$ structure. But this implies there is an extra invariant section $\tilde{\Xi}\in\Gs{X}$ satisfying $\dd_F\tilde{\Xi}=0$ and indeed such a generalised tensor exists: the globally defined section $\dd y^5$ satisfies 
\begin{equation}
    \dd_F (\dd y^5) = \tilde{F}\wedge \dd y^5 
        = - *_\delta \dd H \wedge \dd y^5 = 0 \, ,      
\end{equation}
given we have smeared so that $\partial H/\partial y^5=0$. Rather than $\dd y^5$ itself, the $\dd_F$-closed combination that leads to an additional tensor multiplet is the one orthogonal to $\Xi^5$, namely $\tilde{\Xi}= 2\dd y^5 -\Xi^5=\dd y^5 - H\, {*_\delta \dd y^5}$. 

This pattern extends to all the other brane consistent truncations: if we smear the solution over $k$ directions\footnote{If $k=n$ the flux $\tilde{F}$ vanishes, and the geometry is simply that of a flat torus with no brane. There is a torsion-free identity structure and one  gets the standard maximally supersymmetric consistent truncation.}  (with $k<n$), then we get a torsion-free $Spin(n-k)\subset Spin(n,n)$ structure and $k$ additional $\dd_F$-closed singlets of the form $\tilde{\Xi}^i=\dd y^i - (-1)^{[n/2]}H\, {*_\delta \dd y^i}$ where $i=n-k+1,\dots,n$. The corresponding truncated theory is then half-maximal supergravity coupled to $k$ additional vector multiplets in the case of the D$p$-branes and IIB NS5-brane, and $k$ tensor multiplets in the case of the M5-brane. This can be equally well understood as first making a maximally supersymmetric consistent truncation on $T^k$ to a $(10-k)$- or $(11-k)$-dimensional theory and then considering the truncation on the brane solution in that reduced theory. 
 
Finally, it is perhaps worth emphasising that the class of consistent truncations we consider here, although comparatively simple, is one of the few known cases where the flux, or equivalently the exceptional generalised geometry, was essential to understanding the truncation. Although the background admits a conventional identity structure, that is, is parallelisable, this structure does not have singlet torsion and so cannot explain the consistent truncation. It is only once one goes to the generalised geometry that one finds a torsion-free structure defined by the generalised $\Xi^m$ tensors. This relates to an interesting classification problem. For maximally supersymmetric truncations with compact gauge groups, the sphere and group manifold cases give all the possible consistent truncations.\footnote{The analysis of the maximally symmetric case reduces to an algebraic problem~\cite{Inverso:2017lrz,Bugden:2021wxg,Hulik:2023aks}. The internal manifold is a coset $G/K$ and if $G$ is compact, compatibility with the gauged supergravity embedding tensor conditions (see for example~\cite{Trigiante:2016mnt}) already restricts to the sphere and group manifold cases. If $G$ is noncompact, and hence not necessarily reductive, the general classification remains open.}. For the half-maximal case the corresponding classification remains an open question. We do not know many intrinsically generalised geometry examples there are beyond those based on simple brane solutions discussed here.

\section*{Acknowledgments}
We are grateful to Charles Strickland--Constable for helpful discussions. The work of KSS and DW is supported in part by the STFC Consolidated Grant ST/X000575/1. DW also thanks the Galileo Galilei Institute for Theoretical Physics and the Leinweber Institute for Theoretical Physics at UC Berkeley for hospitality and support during the completion of this work. 
JL thanks the Center for Joint Quantum Studies and Department of Physics at Tianjin University for hospitality and support during the completion of this work.

\newpage

\begin{appendices}
\section{Conventions}\label{app:conventions}
For clarity, we outline our convention throughout the paper in this Appendix. 
\begin{itemize}
    \item We denote ten- or eleven-dimensional quantities with hats, and lower-dimensional quantities without hats.
    \item We denote fields in the $Spin(n,n)\times\mathbb{R}^+$ background with tildes, \ie\ $\{\tilde{g}_{mn},\tilde{A}_{m_1\cdots m_{n-2}},\tilde{\Delta}\}$.
    \item We distinguish bundles in exceptional generalised geometry by hats, while those in $Spin(n,n)\times\mathbb{R}^+$ generalised geometry are written without hats.
    \item We denote indices on the exceptional generalised tangent bundle by $M,N,\cdots$, those on the external space by $\mu,\nu,\cdots$, and those on the internal space by $m,n,\cdots$.
\end{itemize}

\section{Exceptional generalised geometry}\label{app:egg}
Exceptional generalised geometry is the study of structures on a generalised
tangent bundle $\hat{E}$, where $\hat{E}$ admits a unique action of the $E_{n(n)}$ group \cite{Hull:2007zu, PiresPacheco:2008qik}. In this section, we briefly summarise the essential results of exceptional generalised geometry for M-theory and type IIB, following \cite{Coimbra:2011ky,Ashmore:2015joa}.  The type IIA results can be obtained by dimensional reduction from M-theory. A comprehensive construction of type IIA exceptional generalised geometry is given in \cite{Cassani:2016ncu}

For $n\leq 7$, M-theory on an $n$-dimensional manifold $M$ has ageneralised tangent bundle of the form 
\begin{equation}
    \hat{E}\simeq TM\oplus \Lambda^2T^*M \oplus \Lambda^5T^*M \oplus (T^*M\otimes\Lambda^7T^*M)\,,
\end{equation}
while for type IIB on a $(n-1)$-dimensional manifold $M$, the generalised tangent bundle is
\begin{equation}
    \hat{E}\simeq TM\oplus (S\otimes T^*M) \oplus \Lambda^3T^*M \oplus (S\otimes \Lambda^5T^*M)\oplus (T^*M\otimes\Lambda^6T^*M)\,,
\end{equation}
where $S$ is an $\mathbb{R}^2$ bundle transforming as a doublet of $SL(2,\mathbb{R})$. In both cases, the generalised tangent bundle transforms in the vector representation of  $E_{n(n)}\times\mathbb{R}^+$ with $\mathbb{R}^+$ weight one. A weight $p$ scalar field is a section of $\det(T^*M)^{p/(n-9)}$. 

The generalised frame bundle $\tilde{F}$ is an $E_{n(n)}\times \mathbb{R}^+$ principal bundle constructed from all possible generalised frames of $\hat{E}$. One defines generalised tensors as sections of the vector bundles associated with different $E_{n(n)}\times\mathbb{R}^+$ representations. Of particular interest is the adjoint bundle $\widehat{\mathrm{ad}}$. In M-theory, we have 
\begin{equation}
    \begin{aligned}
        \widehat{\mathrm{ad}}\simeq\mathbb{R} \oplus (TM\otimes T^*M)\oplus \Lambda^3T^*M\oplus \Lambda^6T^*M\oplus \Lambda^3 TM\oplus\Lambda^6T^M\,,
    \end{aligned}
\end{equation}
while in type IIB 
\begin{equation}
    \begin{aligned}
        \widehat{\mathrm{ad}} &\simeq\mathbb{R}\oplus (TM\otimes T^*M) \oplus (S\otimes S^*)_0 \oplus (S\otimes\Lambda^2TM) \oplus (S\otimes\Lambda^2T^*M)\\
        &\qquad \qquad {} \oplus \Lambda^4TM\oplus\Lambda^4T^*M\oplus (S\otimes\Lambda^6TM) \oplus (S\otimes\Lambda^6T^*M)
    \end{aligned}
\end{equation}
where the subscript on $(S\otimes S^*)_0$ donates the traceless part.

Each of these bundles are naturally defined as an extension, so that there is non-trivial patching between their tensor components. For example, on the overlap of two patches $U_i \cap U_j$ in M-theory, a generalised vector $V\in\Gamma(\hat{E})$ is patched by 
\begin{equation}
    V_{(i)}=e^{\dd\Lambda_{ij}+\dd\tilde\Lambda_{ij}}\,V_{(j)}\,.
\end{equation}
Here, $\Lambda_{ij}$ and $\tilde\Lambda_{ij}$ are locally defined two- and five-forms in M-theory, and one- and three-forms in type IIB, respectively. Their exterior derivatives are viewed as sections of $\mathrm{ad}\tilde{F}$ so that $e^{\dd\Lambda_{ij}+\dd\tilde\Lambda_{ij}}$ acts via the exponentiated adjoint action. The patching of each generalised tensor bundle then follows by the using the corresponding adjoint action. These are given explicitly in, for example, \cite{Coimbra:2011ky,Ashmore:2015joa}

There is a further bundle $\hat{N}$ that plays an important role in consistent truncations. In M-theory, it is given by 
\begin{equation}
    \begin{aligned}
        \hat{N} &\simeq T^*M\oplus \Lambda^4 T^*M \oplus (T^*M\otimes\Lambda^6 T^*M) \\
        & \qquad  {} \oplus (\Lambda^3T^*M\otimes\Lambda^7T^*M)\oplus (\Lambda^6T^*M\otimes\Lambda^7T^*M)\,.
    \end{aligned}
\end{equation}
while in type IIB, it has the form 
\begin{equation}
\begin{aligned}
    \hat{N} &\simeq S \oplus \Lambda^2T^*M\oplus (S\otimes \Lambda^4T^*M) 
       \oplus (T^*M\otimes\Lambda^5 T^*M) \\ & \qquad {}
       \oplus (S\otimes S^*\otimes \Lambda^6 T^*M)
       \oplus (S\otimes \Lambda^2 T^*M \otimes \Lambda^6 T^*M) \\& \qquad {}
       \oplus (\Lambda^4 T^*M \otimes \Lambda^6 T^*M)
       \oplus (S\otimes (\Lambda^6 T^*M)^{\otimes2}) \,.
\end{aligned}
\end{equation}

The isomorphism for all these generalised bundles are non-unique and depend on a choice of form-field potentials. In the M-theory case, given potentials $A\in\Gamma(\Lambda^3T^*M)$ and $\tilde{A}\in\Gamma(\Lambda^6T^*M)$ that are patched by
\begin{equation}
    A_{(i)} = A_{(j)} +\dd\Lambda_{(ij)}\,,\qquad \tilde{A}_{(i)}=\tilde{A}_{(j)}+\dd\tilde{\Lambda}_{(ij)} - \frac12 \dd\Lambda_{(ij)}\wedge A_{(j)}\,,
\end{equation} 
one can construct twisted sections of $\hat{E}$ and $\widehat{\mathrm{ad}}$ via
\begin{equation}
    V=e^{A+\tilde{A}}\,\tilde{V}\,,\qquad R=e^{A+\tilde{A}}\,\tilde{R}\,e^{-A-\tilde{A}}\,,
\end{equation}
where $V\in\Gamma(\hat{E})$ and $R\in\Gamma(\mathrm{ad}\tilde{F})$, while the ``untwisted'' objects $\tilde{V}$ and $\tilde{R}$ are sections of $TM\oplus \Lambda^2T^*M \oplus \cdots$ and $\mathbb{R} \oplus (TM\otimes T^*M)\oplus\cdots$, respectively. The corresponding field strengths of the potentials 
\begin{equation}
    F=\dd A\,,\qquad \tilde{F}=\dd\tilde{A} -\frac12 A\wedge F
\end{equation}
are precisely the corresponding supergravity field strengths. Similarly, in the type IIB case, the twisted sections of $\hat{E}$ and $\widehat{\mathrm{ad}}$ are constructed as 
\begin{equation}
    V=e^{B^i+C}\,\tilde{V}\,,\qquad R=e^{B^i+C}\,\tilde{R}\,e^{-B^i-C}\,.
\end{equation}
The corresponding field strengths are 
\begin{equation}
    F^i =\dd B^i \,,\qquad F=\dd C+\frac{1}{2}\epsilon_{ij}B^i\wedge F^j\,,
\end{equation}
where $F^1=H_3,\ F^2=F_3$ and $F$ are the corresponding supergravity field strengths. One can construct the twisted section of $\hat{N}$ bundle similarly by considering it as a sub-bundle of $S^2\hat{E}$.

Crucially the generalised tangent space defines a generalised Lie derivative, also known as the Dorfman derivative, which encodes the bosonic symmetries of the supergravity. In M-theory, the Dorfman derivative along a generalised vector $V=v+\omega+\sigma+\tau\in\Gamma(\hat{E})$, acting on any generalised tensor, takes the form 
\begin{equation}
\label{eq:Dorfman-derivative-def-M}
    L_V = \mathcal{L}_v - (\dd \omega + \dd \sigma)\, \cdot
\end{equation}
where $\mathcal{L}_v$ is the ordinary Lie derivative and the second term acts via the adjoint action. In IIB, given $V=v+\lambda^i+\rho+\sigma\in\Gamma(\hat{E})$ we have 
\begin{equation}
\label{eq:Dorfman-derivative-def-IIB}
L_V = \mathcal{L}_v - (\dd \lambda^i+ \dd \rho + \dd\sigma^i)\, \cdot 
\end{equation}

\section{$Spin(n,n)\times\mathbb{R}^+$ generalised geometry}\label{app:sgg}
We provide details of the construction $Spin(n,n)\times\mathbb{R}^+$ using the $GL(n)$ subgroup, including useful representations and tensor products.

The $Spin(n,n)\times\mathbb{R}^+$ generalised geometry was constructed  by introducing the bundle \cite{Strickland-Constable:2013xta}
\begin{equation}
    W=\left(\Lambda^{n}T^{*}M\right)^{1/2}\otimes TM\,.
    \label{eq:identify-WT}
\end{equation}
By constrast, for $SO(n,n)\times\mathbb{R}^+$ generalised geometry one takes $W=TM$. The generalised tangent bundle 
\begin{equation}
    E\simeq TM\oplus\Lambda^{n-3}T^{*}M\oplus\left(\Lambda^{n}T^{*}M\otimes\Lambda^{n-5}T^{*}M\right)\oplus\left(\left(\Lambda^{n}T^{*}M\right)^{\otimes2}\otimes\Lambda^{n-7}T^{*}M\right)\oplus\cdots\,.
\end{equation}
is in the spinor representation of $Spin(n, n)$ with the chirality depending on the $n\,.$ For odd $n$, the generalised tangent bundle $E$ has positive chirality and is realised as
\begin{equation}
    S_{(n-4)/2}^{+}\simeq\left(\Lambda^{n}W\right)^{1/2}\otimes\Lambda^{\mathrm{even}}W^{*}\otimes\left(\Lambda^{n}T^{*}\right)^{\frac{n-4}{4}} \,.
\end{equation}
whereas for even $n$ it has negative chirality and is constructed from 
\begin{equation}
    S_{(n-4)/2}^{-}\simeq\left(\Lambda^{n}W\right)^{1/2}\otimes\Lambda^{\mathrm{odd}}W^{*}\otimes\left(\Lambda^{n}T^{*}\right)^{\frac{n-4}{4}} \,.
\end{equation}
The subscript on $S_{(n-4)/4}^{\pm}$ denotes the $\mathbb{R}^+$ weight, with 
\begin{equation}
    \mathbf{1}_{1}\simeq \left(\Lambda^{n}T^{*}\right)^{1/2} \,.
    \label{eq:R-weighted}
\end{equation}
We write the section of $E$ bundle as
\begin{equation}
    V=v+\lambda+\rho+\sigma+\cdots,
\end{equation}
where $v\in\Gamma(TM)\,,\ \lambda\in\Gamma(\Lambda^{n-3}T^*M)\,,\ \rho \in \Gamma\left(\Lambda^{n}T^{*}M\otimes\Lambda^{n-5}T^{*}M\right)$ and so on.

On the other hand, the opposite-chirality generalised bundle 
\begin{equation}
    \begin{aligned}
    P\simeq\Lambda^{n}T^{*}M\oplus\Lambda^{2}T^{*}M
    \oplus
    \Lambda^{n-4}TM\oplus\left(\Lambda^{n}TM\otimes\Lambda^{n-6}TM\right)
    \oplus\cdots,
    \end{aligned}
\end{equation}
is constructed from 
\begin{equation}
    \begin{aligned}
        \left(S_{(n-6)/2}^{-}\right)^{*}
        &\simeq \left(\Lambda^{n}W^{*}\right)^{1/2}\otimes\Lambda^{\mathrm{odd}}W\otimes\left(\Lambda^{n}T\right)^{\frac{n-6}{4}}
        && \text{for odd } n, \\
        \left(S_{(n-6)/2}^{+}\right)^*
        &\simeq \left(\Lambda^{n}W^*\right)^{1/2}\otimes\Lambda^{\mathrm{even}}W\otimes\left(\Lambda^{n}T\right)^{\frac{n-6}{4}}
        && \text{for even } n.
    \end{aligned}
    \label{eq:bundle-construction-F}
\end{equation}
We write a section of  $P$ as 
\begin{equation}
    U=\omega+\kappa+\mu+\nu+\cdots\,,
\end{equation}
where $\omega\in\Gamma(\Lambda^{n}T^{*}M)\,,\ \kappa\in\Gamma(\Lambda^{2}T^{*}M)\,,\ \mu\in\Gamma\left(\Lambda^{n}TM\otimes\Lambda^{n-6}TM\right)$ and so on.

The adjoint bundle is 
\begin{equation}
    \mathrm{ad}\simeq\mathbb{R}\oplus\left(TM\otimes T^{*}M\right)\oplus\Lambda^{n-2}TM\oplus\Lambda^{n-2}T^{*}M\,.
\end{equation}
We write the section of the adjoint bundle as 
\begin{equation}
    R=l+r+\beta+b\,,
\end{equation}
with $l\in\mathbb{R}\,,\ r\in\Gamma(\text{End}\,TM)\,\ \beta\in\Gamma(\Lambda^{(n-2)}TM)$ and $b\in\Gamma(\Lambda^{(n-2)}T^*M)$. The adjoint action of $R\in\Gamma(\mathrm{ad})$ on $V\in\Gamma(E)$ to be $V'=R\cdot V$ is then given by 
\begin{equation}
    \begin{aligned}
        v'&=\frac{(n-4)}{2}l\,v+r\cdot v-\beta\,\lrcorner\,s\left(\lambda\right)\,,\\
        \lambda'&=\frac{(n-4)}{2}l\,\lambda+r\cdot\lambda+v\,\lrcorner\,b+\beta\,\lrcorner\,\rho\,,\\
        \rho'&=\frac{(n-4)}{2}l\,\rho+r\cdot\rho-j^{n-5}\lambda\wedge s\left(b\right)+\beta\,\lrcorner\,\sigma\,,\\
        \sigma'&=\frac{(n-4)}{2}l\,\sigma+r\cdot\sigma-j^{n-7}\rho\wedge s\left(b\right)\,,\\
        &\cdots
    \end{aligned}
    \label{eq:spindd-algebra-action-V}
\end{equation}
where $s\left(\lambda\right)=\left(-1\right)^{[m/2]}\lambda$ for $\lambda\in\Gamma\left(\Lambda^{m}T^{*}\right)\,$. The factor $\frac{n-4}{2}$ in the scaling factor is the weight of the $E$ bundle under $\mathbb{R}^+$. More terms will appear if one considers $n\geq9\,$. One finds that the $Spin(n,n)$ sub-algebra is then generated by setting $l=\frac12r^a{}_a$. Note that this definition of the action is degenerate when $n=2$, since the $l$ and $r$ actions are no longer independent and one needs to instead define a different, independent action of the scalar $l$. We will ignore this subtlety in the following and simply restrict to $n>2$. 

The generalised bundle $X$, transforms in the vector representation of $Spin(n,n)$ and carrys $\mathbb{R}^+$ weight one:
\begin{equation}
    X\simeq (W \oplus W^*) \otimes (\Lambda^nT^*M)^{1/2} \simeq T^{*}M\oplus\Lambda^{n-1}T^{*}M\,.
\end{equation}
We write the section as 
\begin{equation}
    Y=\eta+\tilde{\eta}\,,
\end{equation}
where $\eta\in\Gamma(T^*M)$ and $\tilde\eta\in\Gamma(\Lambda^{n-1}T^*M)$. Writing $Y'=R\cdot Y$ for the adjoint action of $R\in\Gamma(\mathrm{ad})$ on $Y\in\Gamma\left(X\right)\,$, the components of $Y'$ are then given by 
\begin{equation}
    \begin{aligned}
        \eta'=&\,l\,\eta+r\cdot\eta+\beta\,\lrcorner\,\tilde{\eta}\,,\\
        \tilde{\eta}'=&\,l\,\tilde{\eta}+r\cdot\tilde{\eta}+\eta\wedge s\left(b\right)\,.
        \label{eq:spindd-algebra-action-X}
    \end{aligned}
\end{equation}

From either~\eqref{eq:spindd-algebra-action-V} or~\eqref{eq:spindd-algebra-action-X} one finds the adjoint algebra $R_{12}=[R_1,R_2]$ has the form  
\begin{equation}
    \begin{aligned}
        l_{12}=&\left(-1\right)^{\left[\frac{n-3}{2}\right]}\frac{2}{n-2}\left(\beta_{2}\,\lrcorner\,b_{1}-\beta_{1}\,\lrcorner\,b_{2}\right)\,,\\
        r_{12}=&\left[r_{1},r_{2}\right]+\left(-1\right)^{\left[\frac{n-3}{2}\right]}\left(j\beta_{2}\,\lrcorner\,jb_{1}-j\beta_{1}\,\lrcorner\,jb_{2}\right)\\
        &\qquad-\left(-1\right)^{\left[\frac{n-3}{2}\right]}\frac{2}{n-2}\left(\beta_{2}\,\lrcorner\,b_{1}-\beta_{1}\,\lrcorner\,b_{2}\right)\mathbf{1}\,,\\
        \beta_{12}=&r_{1}\cdot\beta_{2}-r_{2}\cdot\beta_{1}\,,\\
        b_{12}=&r_{1}\cdot b_{2}-r_{2}\cdot b_{1}\,.
    \end{aligned}
    \label{eq:adjoint-commutator-explicit}
\end{equation}

The splittings of these generalised bundles are non-unique and depend on a choice of  potential. Given $A\in\Gamma(\Lambda^{n-2}T^*M)$ that are patched by
\begin{equation}
    A_{(i)} = A_{(j)} +\dd\Lambda_{(ij)}\,,
\end{equation} 
one can construct twisted sections of $E$ and $\mathrm{ad}$ via
\begin{equation}
    V=e^{A}\,\tilde{V}\,,\qquad R=e^{A}\,\tilde{R}\,e^{-A}\,,
\end{equation}
where $V\in\Gamma(E)$ and $R\in\Gamma(\mathrm{ad})$, while the "untwisted" objects $\tilde{V}$ and $\tilde{R}$ are sections of $TM\oplus\Lambda^{n-3}T^{*}M\oplus\cdots$ and $\mathbb{R} \oplus (TM\otimes T^*M)\oplus\cdots$, respectively. The corresponding field strength of the potential
\begin{equation}
    F=\dd A
\end{equation}
is identified to the supergravity field strengths after embedded into $E_{k(k)}\times\mathbb{R}^+$ generalised geometry. It is similar for other generalised bundles.

Finally the action of the generalised Lie (or Dorfman) derivative of a generalised vector $V\in\Gamma(E)$ on any generalised tensor only depends on the vector $v$ and $(n-3)$-form $\lambda$ components and has the form 
\begin{equation}
    L_{V}= \mathcal{L}_v - \dd \lambda \, \cdot 
    \label{eq:Dorfman-derivative-Spin}
\end{equation}
similar to the exceptional generalised geometry expressions~\eqref{eq:Dorfman-derivative-def-M} and~\eqref{eq:Dorfman-derivative-def-IIB}.

\section{The IIA NS5-brane truncation}\label{app:NS5A}
In this section, we give direct proof of the consistency of the truncation ansatz \eqref{eq:NS5A-embedding-ansatz} by reproducing the equations of motion of 6d $(2,0)$ supergravity with an additional tensor multiplet \cite{Romans:1986er, Riccioni:1997np} from type IIA equations (provided in \cite{Lin:2025ucv}\footnote{We change the convention here as $\hat{F}_4^{\text{here}} = \tilde{F}_4^{LSS}$}.) while keeping unknown constants $\alpha$, $\beta$ and $\gamma$ and functions $a(y)$ and $b(y)$ as
\begin{equation}
    \begin{aligned}
        &\dd \hat{s}_E^2=H^{-\frac14}\ee^{\alpha\sigma}g_{\mu\nu}\dd x^\mu \dd x^\nu + H^{\frac34}\ee^{\beta\sigma}\delta_{mn}\dd y^m \dd y^n\,,\\ 
        &\hat{H}_3=-*_\delta\dd H + H^5_3\,,\qquad \hat{F}_2=a\,\dd \phi ^m\wedge\dd y^m\,,\qquad \ee^{\hat{\Phi}} = H^{\frac12}e^{\gamma\sigma}\,,\\
        &\hat{F}_4=H^m_3\wedge\dd y^m+b\,\dd\phi^m\wedge *_\delta \dd y^m\,.
    \end{aligned}
    \label{eq:NS5A-ansatz-unknown}
\end{equation}

Before proceeding, it is useful to record the identities
\begin{equation}
    \begin{aligned}
        *\hat{F}_{2}=&-H^{\frac{1}{4}}e^{\left(2\alpha+\beta\right)\sigma}a*_{g}\left(\dd\phi^{m}\right)\wedge*_{\delta}\dd y^{m}\,.\\
        *\hat{H}_{3}=&H^{-\frac{3}{2}}e^{\left(3\alpha-\beta\right)\sigma}\mathrm{vol}_{g}\wedge\dd H+H^{\frac{3}{2}}e^{2\beta\sigma}*_{g}H_{3}^{5}\wedge\mathrm{vol}_{\delta}\,.\\
        *\hat{F}_{4}=&-H^{\frac{3}{4}}e^{\beta\sigma}*_{g}H_{3}^{m}\wedge*_{\delta}\dd y^{m}+H^{-\frac{5}{4}}e^{\left(2\alpha-\beta\right)\sigma}b*_{g}\left(\dd\phi^{m}\right)\wedge\dd y^{m}\,.
    \end{aligned}
\end{equation}
Under the ansatz \eqref{eq:NS5A-ansatz-unknown}, the $\hat{F}_2$ equation of motion reduces to  
\begin{equation}
    \begin{aligned}
        &a=H^{-1}e^{\left(-2\beta-\gamma\right)\sigma}b\,,\\
        &\dd\left(e^{\left(2\alpha+\beta+\frac{3}{2}\gamma\right)\sigma}a*_{g}\dd\phi^{m}\right)=-e^{\left(\beta+\frac{\gamma}{2}\right)\sigma}H_{3}^{5}\wedge*_{g}H_{3}^{m}\,.
    \end{aligned}
\end{equation}
Since there is no $y$-dependence on the right-hand side of the second equation, $a$ must also be independent of $y$, and hence is a constant. The first equation then implies
\begin{equation}
    2\beta+\gamma = 0\,.
    \label{eq:para-1}
\end{equation}
For later convenience, we introduce the constant 
\begin{equation}
    \tilde{b}=H^{-1}b=a\,.
\end{equation}
Then, under the ansatz \eqref{eq:NS5A-ansatz-unknown}, the $\hat{F}_4$ equation of motion reduces to
\begin{equation}
    \begin{aligned}
        &e^{\left(\beta+\frac{1}{2}\gamma\right)\sigma}*_{g}H_{3}^{m}=H_{3}^{m}\,,\\
        &\dd\left(e^{\left(\beta+\frac{1}{2}\gamma\right)\sigma}*_{g}H_{3}^{m}\right)=\tilde{b}H_{3}^{5}\wedge\dd\phi^{m}\,,\\
        &\dd\left(e^{\left(2\alpha-\beta+\frac{1}{2}\gamma\right)\sigma}\tilde{b}*_{g}\dd\phi^{m}\right)=-H_{3}^{5}\wedge H_{3}^{m}\,.
    \end{aligned}
\end{equation}
Using \eqref{eq:para-1}, the first equation is precisely the self-duality condition of $H^m_3$. 
Likewise, under the ansatz \eqref{eq:NS5A-embedding-ansatz} and the self-duality condition of $H^m_3$, the $\hat{H}_3$ equation of motion reduces to
\begin{equation}
    \dd\left(e^{\left(2\beta-\gamma\right)\sigma}*_{g}H_{3}^{5}\right)=-e^{\left(-\beta-\frac{1}{2}\gamma\right)\sigma}\tilde{b}*_{g}H_{3}^{m}\wedge\dd\phi^{m}-\tilde{b}H_{3}^{m}\wedge\dd\phi^{m}\,.
\end{equation}
We next consider the Bianchi identities. Firstly, $\dd\hat{F}_{2}=0$ is trivial. Next, $\dd\hat{H}_{3}=0\,$ provides us  
\begin{equation}
    \dd H_{3}^{5}=0\,.
\end{equation}
Moreover, the $\dd\hat{F}_{4}=\hat{F}_{2}\wedge\hat{H}_{3}$ reduces to 
\begin{equation}
    a=\tilde{b}\,,\quad \text{and}\quad \dd H_{3}^{m}=aH_{3}^{5}\wedge\partial_{\mu}\phi^{m}\dd x^{\mu}\,.
\end{equation}
The dilaton equation of motion under the ansatz \eqref{eq:NS5A-embedding-ansatz} reduces to 
\begin{equation}
    -2\beta\dd\left(e^{\left(2\alpha+2\beta\right)\sigma}*_{g}\dd\sigma\right)=e^{\left(2\alpha-2\beta\right)\sigma}a^{2}\dd\phi^{m}\wedge*_{g}\dd\phi^{m}-\frac{1}{2}e^{4\beta\sigma}H_{3}^{5}\wedge*_{g}H_{3}^{5}+\frac{1}{4}H_{3}^{m}\wedge*_{g}H_{3}^{m}\,.
\end{equation}

To analyse the Einstein equation, we first introduce the Zehnbein
\begin{equation}
    e^{\underline{\mu}}=H^{-\frac{1}{8}}e^{\frac{1}{2}\alpha\sigma}\tilde{e}^{\underline{\mu}}\,,\qquad e^{\underline{m}}=H^{\frac{3}{8}}e^{\frac{1}{2}\beta\sigma}\dd y^{m}\,,
    \label{eq:NS5A-scalar-eq-1}
\end{equation}
where $\tilde{e}^{\underline{\mu}}$ is the Sechsbein of the metric $g_{\mu\nu}$. Using the Cartan's structure equations, one obtain the connection 1-form and curvature 2-form and then the Ricci tensor
\begin{equation}
    \begin{aligned}
        R_{\mu\nu}=&\tilde{R}_{\mu\nu}+\left(\alpha^{2}+2\alpha\beta-\beta^{2}\right)\partial_{\mu}\sigma\partial_{\nu}\sigma-\alpha\left(\alpha+\beta\right)\left(g^{\gamma\eta}\partial_{\gamma}\sigma\partial_{\eta}\sigma\right)g_{\mu\nu}\\
        &-2\left(\alpha+\beta\right)\tilde{D}_{\nu}\left(\partial_{\mu}\sigma\right)-\frac{\alpha}{2}\tilde{D}_{\rho}\left(g^{\rho\eta}\partial_{\eta}\sigma\right)g_{\mu\nu}+\frac{1}{8}g_{\mu\nu}H^{-1}e^{\left(\alpha-\beta\right)\sigma}\partial_{m}H^{-1}\partial_{m}H\,,\\
        R_{\mu m}=&-\frac{\beta}{2}H^{-1}\partial_{m}H\partial_{\mu}\sigma\,,\\
        R_{mn}=&-\frac{\beta}{2}He^{-\left(\alpha-\beta\right)\sigma}\tilde{D}_{\rho}\left(g^{\rho\eta}\partial_{\eta}\sigma\right)\delta_{mn}+\frac{3}{8}\partial_{m}H^{-1}\partial_{n}H-\frac{3}{8}\delta_{mn}\partial_{p}H^{-1}\partial_{p}H\\
        &-He^{-\left(\alpha-\beta\right)\sigma}\left(\alpha+\beta\right)\beta\left(g^{\rho\eta}\partial_{\rho}\sigma\partial_{\eta}\sigma\right)\delta_{mn}\,.
    \end{aligned}
\end{equation}
where $\tilde{D}$ is the covariant derivative with respect to the spin connection in $g_{\mu\nu}\,$. Using the scalar equation \eqref{eq:NS5A-scalar-eq-1}, one obtains the only non-trivial equation 
\begin{equation}
    \begin{aligned}
        \tilde{R}_{\mu\nu}=&\left(-\alpha^{2}-2\alpha\beta+3\beta^{2}\right)\partial_{\mu}\sigma\partial_{\nu}\sigma+2\left(\alpha+\beta\right)\tilde{D}_{\nu}\left(\partial_{\mu}\sigma\right)\\&+e^{-4\beta\sigma}a^{2}\left(\partial_{\mu}\phi^{m}\partial_{\nu}\phi^{m}-\frac{1}{4\beta}\left(\alpha+\beta\right)\partial_{\rho}\phi^{m}\partial^{\rho}\phi^{m}g_{\mu\nu}\right)\\&+\frac{1}{4}e^{-2\left(\alpha-\beta\right)\sigma}\left[H_{\mu}^{5\ \rho\sigma}H_{\nu\rho\sigma}^{5}-\frac{1}{2\cdot3!}\left(1-\frac{\alpha}{\beta}\right)H_{\rho\sigma\lambda}^{5}H^{5\rho\sigma\lambda}g_{\mu\nu}\right]\\&+\frac{1}{4}e^{-2\left(\alpha+\beta\right)\sigma}H_{\mu}^{m\rho\sigma}H_{\nu\rho\sigma}^{m}\,.
    \end{aligned}
\end{equation}

Collecting the above results, we obtain
\begin{equation}
    b=H\tilde{b}\,,\qquad a=\tilde{b}\,,\qquad2\beta+\gamma=0\,,
\end{equation}
Bianchi identities 
\begin{equation}
    0=\dd H_{3}^{5}\,,\qquad 0=\dd H_{3}^{m}-aH_{3}^{5}\wedge\dd\phi^{m}\,.
\end{equation}
and equations of motion
\begin{equation}
    \begin{aligned}
        0=&\dd\left(e^{4\beta\sigma}*_{g}H_{3}^{5}\right)+2aH_{3}^{m}\wedge\dd\phi^{m}\,,\\
        0=&\dd\left(e^{2\left(\alpha-\beta\right)\sigma}a*_{g}\dd\phi^{m}\right)+H_{3}^{5}\wedge H_{3}^{m}\,,\\
        0=&\dd\left(e^{2\left(\alpha+\beta\right)\sigma}*_{g}\dd\sigma\right)+\frac{1}{2\beta}a^{2}e^{2\left(\alpha-\beta\right)\sigma}\dd\phi^{m}\wedge*_{g}\dd\phi^{m}-\frac{1}{4\beta}e^{4\beta\sigma}H_{3}^{5}\wedge*_{g}H_{3}^{5}\,,\\
        \tilde{R}_{\mu\nu}=&\left(-\alpha^{2}-2\alpha\beta+3\beta^{2}\right)\partial_{\mu}\sigma\partial_{\nu}\sigma+2\left(\alpha+\beta\right)\tilde{D}_{\nu}\left(\partial_{\mu}\sigma\right)\\
        &+e^{-4\beta\sigma}a^{2}\left(\partial_{\mu}\phi^{m}\partial_{\nu}\phi^{m}-\frac{1}{4\beta}\left(\alpha+\beta\right)\partial_{\rho}\phi^{m}\partial^{\rho}\phi^{m}g_{\mu\nu}\right)\\
        &+\frac{1}{4}e^{-2\left(\alpha-\beta\right)\sigma}\left[H_{\mu}^{5\ \rho\sigma}H_{\nu\rho\sigma}^{5}-\frac{1}{2\cdot3!}\left(1-\frac{\alpha}{\beta}\right)g_{\mu\nu}H_{\rho\sigma\lambda}^{5}H^{5\rho\sigma\lambda}\right]\\
        &+\frac{1}{4}e^{-2\left(\alpha+\beta\right)\sigma}H_{\mu}^{m\rho\sigma}H_{\nu\rho\sigma}^{m}
    \end{aligned}
\end{equation}
with the self-duality condition 
\begin{equation}
    H_{3}^{m}=*_{g}H_{3}^{m}\,.
    \label{eq:NS5A-self-dual}
\end{equation}

Requiring a standard kinetic term for the scalars in the Einstein frame gives
\begin{equation}
    \begin{aligned}
        \begin{cases}
        -\alpha^{2}-2\alpha\beta+3\beta^{2}=\frac{1}{2}\\
        2\left(\alpha+\beta\right)=0\\
        a^{2}=\frac{1}{2}
        \end{cases}&\Rightarrow\begin{cases}
        \alpha=-\beta=\pm\frac{1}{2\sqrt{2}}\\
        a=\pm\frac{1}{\sqrt{2}}
        \end{cases}\,.
    \end{aligned}
\end{equation}
Choosing $\alpha=-\beta=-\frac{1}{2\sqrt{2}}$ and $a=\frac{1}{\sqrt{2}}$, we simplify the Bianchi identities to be 
\begin{equation}
    0=\dd H_{3}^{5}\,,\qquad 0=\dd H_{3}^{m}-\frac{1}{\sqrt{2}}H_{3}^{5}\wedge\dd\phi^{m}\,.
    \label{eq:NS5A-bianchi}
\end{equation}
and the equations of motion to be 
\begin{equation}
    \begin{aligned}
        0=&\dd\left(e^{\sqrt{2}\sigma}*_{g}H_{3}^{5}\right)+\sqrt{2}H_{3}^{m}\wedge\dd\phi^{m}\,,\\0=&\dd\left(e^{-\sqrt{2}\sigma}*_{g}\dd\phi^{m}\right)+\sqrt{2}H_{3}^{5}\wedge H_{3}^{m}\,,\\0=&\dd*_{g}\dd\sigma+\sqrt{2}e^{-\sqrt{2}\sigma}\dd\phi^{m}\wedge*_{g}\dd\phi^{m}-\frac{1}{\sqrt{2}}e^{\sqrt{2}\sigma}H_{3}^{5}\wedge*_{g}H_{3}^{5}\,,\\\tilde{R}_{\mu\nu}=&\frac{1}{2}\partial_{\mu}\sigma\partial_{\nu}\sigma+\frac{1}{2}e^{-\sqrt{2}\sigma}\partial_{\mu}\phi^{m}\partial_{\nu}\phi^{m}+\frac{1}{4}H_{\mu}^{m\rho\sigma}H_{\nu\rho\sigma}^{m}\\&+\frac{1}{4}e^{\sqrt{2}\sigma}\left(H_{\mu}^{5\ \rho\sigma}H_{\nu\rho\sigma}^{5}-\frac{1}{6}g_{\mu\nu}H_{\rho\sigma\lambda}^{5}H^{5\rho\sigma\lambda}\right)
    \end{aligned}
    \label{eq:NS5A-eom}
\end{equation}

\

We can then read off the scalar kinetic terms
\begin{equation}
    \mathcal{L}\supset-\frac{1}{2}g^{\mu\nu}\partial_{\mu}\sigma\partial_{\nu}\sigma-\frac{1}{2}e^{-\sqrt{2}\sigma}g^{\mu\nu}\partial_{\mu}\phi^{m}\partial_{\nu}\phi^{m}\equiv-\frac{1}{2}H_{IJ}\left(\varphi\right)g^{\mu\nu}\partial_{\mu}\varphi^{I}\partial_{\nu}\varphi^{J}\,,
\end{equation}
with the redefinition $\varphi^{0}=\sigma\,,\ \varphi^{m}=\phi^{m}\,,\ m=1,\cdots,4\,.$ The target-space metric is therefore 
\begin{equation}
    \dd s_{\left(\text{tar}\right)}^{2}=H_{IJ}\dd\varphi^{I}\dd\varphi^{J}=\dd\sigma^{2}+e^{-\sqrt{2}\sigma}\delta_{mn}\dd\phi^{m}\dd\phi^{n}\,,
\end{equation}
which, under a coordinate transformation $r=\frac{1}{\sqrt{2}}e^{-\frac{\sigma}{\sqrt{2}}}\,,$ becomes
\begin{equation}
    \dd s_{\left(\text{tar}\right)}^{2}=\frac{2\dd r^{2}}{r^{2}}+2r^{2}\delta_{mn}\dd\phi^{m}\dd\phi^{n}\,.
\end{equation}
This is precisely the hyperbolic space $\mathbb{H}^{5}$ parameterised by the coset $SO\left(1,5\right)/SO\left(5\right)\,$. 

To compare with the convention of \cite{Riccioni:1997np}, we can parameterise the coset space $SO(1,5)/SO(5)$ with the $6\times 6$ matrix
\begin{equation}
\begin{aligned}
\mathcal V_I{}^{\underline A} 
&=
\bigl(v_I{}^i,\;x_I\bigr) \\
&=
\begin{pmatrix}
\delta_m{}^n & \frac{1}{\sqrt2}\phi^m & -\frac{1}{\sqrt2}\phi^m \\[4pt]
\frac{1}{\sqrt2}e^{-\sigma/\sqrt2}\phi_n &
\frac14\!\left[e^{-\sigma/\sqrt2}(2-\phi^2)+2e^{\sigma/\sqrt2}\right] &
\frac14\!\left[e^{-\sigma/\sqrt2}(2+\phi^2)-2e^{\sigma/\sqrt2}\right] \\[4pt]
\frac{1}{\sqrt2}e^{-\sigma/\sqrt2}\phi_n &
\frac14\!\left[e^{-\sigma/\sqrt2}(2-\phi^2)-2e^{\sigma/\sqrt2}\right] &
\frac14\!\left[e^{-\sigma/\sqrt2}(2+\phi^2)+2e^{\sigma/\sqrt2}\right]
\end{pmatrix},
\end{aligned}
\end{equation}
where \(m,n=1,\dots,4\), \(i=1,\dots,5\), and \(I=1,\dots,6\), and
\begin{equation}
\phi^2\equiv \phi^m\phi^m.
\end{equation}
In this parametrisation, \(\mathcal V\)  satisfies
\begin{equation}
\mathcal V^T\eta\,\mathcal V=\eta\,, \quad \mathcal V\,\eta\,\mathcal V^T=\eta\,, \quad \eta=\mathrm{diag}(1,1,1,1,1,-1)\,.
\end{equation}
which imply the orthonormality and completeness relations
\begin{equation}
v^{iI}v_I{}^j=\delta^{ij},
\qquad
x^I x_I=-1,
\qquad
v^{iI}x_I=0,
\qquad
v_I{}^i v_J{}^i - x_I x_J = \eta_{IJ}.
\end{equation}
Here, the $SO(1,5)$ indices $I,J$ are raised and lowered with $\eta_{IJ}$, while the \(SO(5)\) indices $i,j$ and \(m,n,\dots\) are raised and lowered with $\delta_{ij}$ and $\delta_{mn}$, respectively.
One then finds that the scalar kinetic term is
\begin{equation}
-x^I x^J\,(\partial_\mu v_I{}^i)(\partial^\mu v_J{}^i)
=
-\frac12\,\partial_\mu\sigma\,\partial^\mu\sigma
-\frac12\,e^{-\sqrt2\sigma}\,\partial_\mu\phi^m\,\partial^\mu\phi^m.
\end{equation}

We have therefore shown that the embedding ansatz \eqref{eq:NS5A-embedding-ansatz} realises the 6d $(2,0)$ supergravity with a tensor multiplet, which is in the Einstein frame with standard kinematic terms, in the NS5-brane solution of type IIA supergravity.

\end{appendices}

\bibliographystyle{JHEP}
\bibliography{SpinGG_and_CT}

\providecommand{\href}[2]{#2}\begingroup\raggedright\begin{thebibliography}{10}

\bibitem{Duff:1984hn}
M.J.~Duff, B.E.W.~Nilsson, C.N.~Pope and N.P.~Warner, \emph{{On the Consistency of the {Kaluza-Klein} Ansatz}}, \href{https://doi.org/10.1016/0370-2693(84)91558-2}{\emph{Phys. Lett. B} {\bfseries 149} (1984) 90}.

\bibitem{Cveti__2000}
M.~Cveti\v{c}, H.~L\"u and C.N.~Pope, \emph{{Consistent Kaluza-Klein sphere reductions}}, \href{https://doi.org/10.1103/physrevd.62.064028}{\emph{Phys. Rev. D} {\bfseries 62} (2000) }.

\bibitem{Lu:2006dh}
H.~Lu, C.N.~Pope and K.S.~Stelle, \emph{{Consistent Pauli sphere reductions and the action}}, \href{https://doi.org/10.1016/j.nuclphysb.2007.05.017}{\emph{Nucl. Phys. B} {\bfseries 782} (2007) 79} [\href{https://arxiv.org/abs/hep-th/0611299}{{\ttfamily hep-th/0611299}}].

\bibitem{Scherk:1978ta}
J.~Scherk and J.H.~Schwarz, \emph{{Spontaneous Breaking of Supersymmetry Through Dimensional Reduction}}, \href{https://doi.org/10.1016/0370-2693(79)90425-8}{\emph{Phys. Lett. B} {\bfseries 82} (1979) 60}.

\bibitem{Scherk:1979zr}
J.~Scherk and J.H.~Schwarz, \emph{{How to Get Masses from Extra Dimensions}}, \href{https://doi.org/10.1016/0550-3213(79)90592-3}{\emph{Nucl. Phys. B} {\bfseries 153} (1979) 61}.

\bibitem{deWit:1986oxb}
B.~de~Wit and H.~Nicolai, \emph{{The Consistency of the $S^7$ Truncation in D=11 Supergravity}}, \href{https://doi.org/10.1016/0550-3213(87)90253-7}{\emph{Nucl. Phys. B} {\bfseries 281} (1987) 211}.

\bibitem{Nastase:1999cb}
H.~Nastase, D.~Vaman and P.~van Nieuwenhuizen, \emph{{Consistent nonlinear K K reduction of 11-d supergravity on $AdS(7) \times S(4)$ and selfduality in odd dimensions}}, \href{https://doi.org/10.1016/S0370-2693(99)01266-6}{\emph{Phys. Lett. B} {\bfseries 469} (1999) 96} [\href{https://arxiv.org/abs/hep-th/9905075}{{\ttfamily hep-th/9905075}}].

\bibitem{Nastase:1999kf}
H.~Nastase, D.~Vaman and P.~van Nieuwenhuizen, \emph{{Consistency of the $AdS(7) \times S(4)$ reduction and the origin of selfduality in odd dimensions}}, \href{https://doi.org/10.1016/S0550-3213(00)00193-0}{\emph{Nucl. Phys. B} {\bfseries 581} (2000) 179} [\href{https://arxiv.org/abs/hep-th/9911238}{{\ttfamily hep-th/9911238}}].

\bibitem{Cvetic:2000nc}
M.~Cvetic, H.~Lu, C.N.~Pope, A.~Sadrzadeh and T.A.~Tran, \emph{{Consistent SO(6) reduction of type IIB supergravity on $S^5$}}, \href{https://doi.org/10.1016/S0550-3213(00)00372-2}{\emph{Nucl. Phys. B} {\bfseries 586} (2000) 275} [\href{https://arxiv.org/abs/hep-th/0003103}{{\ttfamily hep-th/0003103}}].

\bibitem{Lee:2014mla}
K.~Lee, C.~Strickland-Constable and D.~Waldram, \emph{{Spheres, generalised parallelisability and consistent truncations}}, \href{https://doi.org/10.1002/prop.201700048}{\emph{Fortsch. Phys.} {\bfseries 65} (2017) 1700048} [\href{https://arxiv.org/abs/1401.3360}{{\ttfamily 1401.3360}}].

\bibitem{Baguet:2015sma}
A.~Baguet, O.~Hohm and H.~Samtleben, \emph{{Consistent Type IIB Reductions to Maximal 5D Supergravity}}, \href{https://doi.org/10.1103/PhysRevD.92.065004}{\emph{Phys. Rev. D} {\bfseries 92} (2015) 065004} [\href{https://arxiv.org/abs/1506.01385}{{\ttfamily 1506.01385}}].

\bibitem{Gauntlett:2007ma}
J.P.~Gauntlett and O.~Varela, \emph{{Consistent Kaluza-Klein reductions for general supersymmetric AdS solutions}}, \href{https://doi.org/10.1103/PhysRevD.76.126007}{\emph{Phys. Rev. D} {\bfseries 76} (2007) 126007} [\href{https://arxiv.org/abs/0707.2315}{{\ttfamily 0707.2315}}].

\bibitem{Kashani-Poor:2007nby}
A.-K.~Kashani-Poor, \emph{{Nearly Kaehler Reduction}}, \href{https://doi.org/10.1088/1126-6708/2007/11/026}{\emph{JHEP} {\bfseries 11} (2007) 026} [\href{https://arxiv.org/abs/0709.4482}{{\ttfamily 0709.4482}}].

\bibitem{Gauntlett:2007sm}
J.P.~Gauntlett and O.~Varela, \emph{{D=5 SU(2) x U(1) Gauged Supergravity from D=11 Supergravity}}, \href{https://doi.org/10.1088/1126-6708/2008/02/083}{\emph{JHEP} {\bfseries 02} (2008) 083} [\href{https://arxiv.org/abs/0712.3560}{{\ttfamily 0712.3560}}].

\bibitem{Gauntlett:2009zw}
J.P.~Gauntlett, S.~Kim, O.~Varela and D.~Waldram, \emph{{Consistent supersymmetric Kaluza-Klein truncations with massive modes}}, \href{https://doi.org/10.1088/1126-6708/2009/04/102}{\emph{JHEP} {\bfseries 04} (2009) 102} [\href{https://arxiv.org/abs/0901.0676}{{\ttfamily 0901.0676}}].

\bibitem{Cassani:2009ck}
D.~Cassani and A.-K.~Kashani-Poor, \emph{{Exploiting N=2 in consistent coset reductions of type IIA}}, \href{https://doi.org/10.1016/j.nuclphysb.2009.03.011}{\emph{Nucl. Phys. B} {\bfseries 817} (2009) 25} [\href{https://arxiv.org/abs/0901.4251}{{\ttfamily 0901.4251}}].

\bibitem{Cassani:2010uw}
D.~Cassani, G.~Dall'Agata and A.F.~Faedo, \emph{{Type IIB supergravity on squashed Sasaki-Einstein manifolds}}, \href{https://doi.org/10.1007/JHEP05(2010)094}{\emph{JHEP} {\bfseries 05} (2010) 094} [\href{https://arxiv.org/abs/1003.4283}{{\ttfamily 1003.4283}}].

\bibitem{Liu:2010sa}
J.T.~Liu, P.~Szepietowski and Z.~Zhao, \emph{{Consistent massive truncations of IIB supergravity on Sasaki-Einstein manifolds}}, \href{https://doi.org/10.1103/PhysRevD.81.124028}{\emph{Phys. Rev. D} {\bfseries 81} (2010) 124028} [\href{https://arxiv.org/abs/1003.5374}{{\ttfamily 1003.5374}}].

\bibitem{Gauntlett:2010vu}
J.P.~Gauntlett and O.~Varela, \emph{{Universal Kaluza-Klein reductions of type IIB to N=4 supergravity in five dimensions}}, \href{https://doi.org/10.1007/JHEP06(2010)081}{\emph{JHEP} {\bfseries 06} (2010) 081} [\href{https://arxiv.org/abs/1003.5642}{{\ttfamily 1003.5642}}].

\bibitem{Skenderis:2010vz}
K.~Skenderis, M.~Taylor and D.~Tsimpis, \emph{{A Consistent truncation of IIB supergravity on manifolds admitting a Sasaki-Einstein structure}}, \href{https://doi.org/10.1007/JHEP06(2010)025}{\emph{JHEP} {\bfseries 06} (2010) 025} [\href{https://arxiv.org/abs/1003.5657}{{\ttfamily 1003.5657}}].

\bibitem{Pope:1987ad}
C.N.~Pope and K.S.~Stelle, \emph{{Zilch Currents, Supersymmetry and {Kaluza-Klein} Consistency}}, \href{https://doi.org/10.1016/0370-2693(87)91487-0}{\emph{Phys. Lett. B} {\bfseries 198} (1987) 151}.

\bibitem{Terrisse:2019usq}
R.~Terrisse and D.~Tsimpis, \emph{{Consistent truncation and de Sitter space from gravitational instantons}}, \href{https://doi.org/10.1007/JHEP07(2019)034}{\emph{JHEP} {\bfseries 07} (2019) 034} [\href{https://arxiv.org/abs/1903.10504}{{\ttfamily 1903.10504}}].

\bibitem{Tsimpis:2020ysl}
D.~Tsimpis, \emph{{Consistent truncation on Calabi-Yau and Nearly-K{\"a}hler manifolds}}, \href{https://doi.org/10.22323/1.376.0122}{\emph{PoS} {\bfseries CORFU2019} (2020) 122} [\href{https://arxiv.org/abs/2002.09359}{{\ttfamily 2002.09359}}].

\bibitem{Lin:2024eqq}
J.~Lin, T.~Skrzypek and K.S.~Stelle, \emph{{Compactification on Calabi-Yau threefolds: consistent truncation to pure supergravity}}, \href{https://doi.org/10.1007/JHEP03(2025)200}{\emph{JHEP} {\bfseries 03} (2025) 200} [\href{https://arxiv.org/abs/2412.00186}{{\ttfamily 2412.00186}}].

\bibitem{Leung:2022nhy}
R.~Leung and K.S.~Stelle, \emph{{Supergravities on branes}}, \href{https://doi.org/10.1007/JHEP09(2022)099}{\emph{JHEP} {\bfseries 09} (2022) 099} [\href{https://arxiv.org/abs/2205.13551}{{\ttfamily 2205.13551}}].

\bibitem{Lin:2025ucv}
J.~Lin, T.~Skrzypek and K.S.~Stelle, \emph{{Interplay of Intersecting Branes and Consistent Embeddings}},  \href{https://arxiv.org/abs/2508.00987}{{\ttfamily 2508.00987}}.

\bibitem{Cassani:2019vcl}
D.~Cassani, G.~Josse, M.~Petrini and D.~Waldram, \emph{{Systematics of consistent truncations from generalised geometry}}, \href{https://doi.org/10.1007/JHEP11(2019)017}{\emph{JHEP} {\bfseries 11} (2019) 017} [\href{https://arxiv.org/abs/1907.06730}{{\ttfamily 1907.06730}}].

\bibitem{Hohm:2014qga}
O.~Hohm and H.~Samtleben, \emph{{Consistent Kaluza-Klein Truncations via Exceptional Field Theory}}, \href{https://doi.org/10.1007/JHEP01(2015)131}{\emph{JHEP} {\bfseries 01} (2015) 131} [\href{https://arxiv.org/abs/1410.8145}{{\ttfamily 1410.8145}}].

\bibitem{Lee:2015xga}
K.~Lee, C.~Strickland-Constable and D.~Waldram, \emph{{New Gaugings and Non-Geometry}}, \href{https://doi.org/10.1002/prop.201700049}{\emph{Fortsch. Phys.} {\bfseries 65} (2017) 1700049} [\href{https://arxiv.org/abs/1506.03457}{{\ttfamily 1506.03457}}].

\bibitem{Malek:2015hma}
E.~Malek and H.~Samtleben, \emph{{Dualising consistent IIA/IIB truncations}}, \href{https://doi.org/10.1007/JHEP12(2015)029}{\emph{JHEP} {\bfseries 12} (2015) 029} [\href{https://arxiv.org/abs/1510.03433}{{\ttfamily 1510.03433}}].

\bibitem{Baguet:2015iou}
A.~Baguet, C.N.~Pope and H.~Samtleben, \emph{{Consistent Pauli reduction on group manifolds}}, \href{https://doi.org/10.1016/j.physletb.2015.11.062}{\emph{Phys. Lett. B} {\bfseries 752} (2016) 278} [\href{https://arxiv.org/abs/1510.08926}{{\ttfamily 1510.08926}}].

\bibitem{Ciceri:2016dmd}
F.~Ciceri, A.~Guarino and G.~Inverso, \emph{{The exceptional story of massive IIA supergravity}}, \href{https://doi.org/10.1007/JHEP08(2016)154}{\emph{JHEP} {\bfseries 08} (2016) 154} [\href{https://arxiv.org/abs/1604.08602}{{\ttfamily 1604.08602}}].

\bibitem{Cassani:2016ncu}
D.~Cassani, O.~de~Felice, M.~Petrini, C.~Strickland-Constable and D.~Waldram, \emph{{Exceptional generalised geometry for massive IIA and consistent reductions}}, \href{https://doi.org/10.1007/JHEP08(2016)074}{\emph{JHEP} {\bfseries 08} (2016) 074} [\href{https://arxiv.org/abs/1605.00563}{{\ttfamily 1605.00563}}].

\bibitem{Malek:2016bpu}
E.~Malek, \emph{{7-dimensional ${\cal N}=2$ Consistent Truncations using $\mathrm{SL}(5)$ Exceptional Field Theory}}, \href{https://doi.org/10.1007/JHEP06(2017)026}{\emph{JHEP} {\bfseries 06} (2017) 026} [\href{https://arxiv.org/abs/1612.01692}{{\ttfamily 1612.01692}}].

\bibitem{duBosque:2017dfc}
P.~du~Bosque, F.~Hassler and D.~L{\"u}st, \emph{{Generalized parallelizable spaces from exceptional field theory}}, \href{https://doi.org/10.1007/JHEP01(2018)117}{\emph{JHEP} {\bfseries 01} (2018) 117} [\href{https://arxiv.org/abs/1705.09304}{{\ttfamily 1705.09304}}].

\bibitem{Malek:2017njj}
E.~Malek, \emph{{Half-Maximal Supersymmetry from Exceptional Field Theory}}, \href{https://doi.org/10.1002/prop.201700061}{\emph{Fortsch. Phys.} {\bfseries 65} (2017) 1700061} [\href{https://arxiv.org/abs/1707.00714}{{\ttfamily 1707.00714}}].

\bibitem{Inverso:2017lrz}
G.~Inverso, \emph{{Generalised Scherk-Schwarz reductions from gauged supergravity}}, \href{https://doi.org/10.1007/JHEP12(2017)124}{\emph{JHEP} {\bfseries 12} (2017) 124} [\href{https://arxiv.org/abs/1708.02589}{{\ttfamily 1708.02589}}].

\bibitem{Malek:2017cle}
E.~Malek and H.~Samtleben, \emph{{Ten-dimensional origin of Minkowski vacua in $N$=8 supergravity}}, \href{https://doi.org/10.1016/j.physletb.2017.11.011}{\emph{Phys. Lett. B} {\bfseries 776} (2018) 64} [\href{https://arxiv.org/abs/1710.02163}{{\ttfamily 1710.02163}}].

\bibitem{Malek:2018zcz}
E.~Malek, H.~Samtleben and V.~Vall~Camell, \emph{{Supersymmetric AdS$_{7}$ and AdS$_6$ vacua and their minimal consistent truncations from exceptional field theory}}, \href{https://doi.org/10.1016/j.physletb.2018.09.037}{\emph{Phys. Lett. B} {\bfseries 786} (2018) 171} [\href{https://arxiv.org/abs/1808.05597}{{\ttfamily 1808.05597}}].

\bibitem{Malek:2019ucd}
E.~Malek, H.~Samtleben and V.~Vall~Camell, \emph{{Supersymmetric AdS$_7$ and AdS$_6$ vacua and their consistent truncations with vector multiplets}}, \href{https://doi.org/10.1007/JHEP04(2019)088}{\emph{JHEP} {\bfseries 04} (2019) 088} [\href{https://arxiv.org/abs/1901.11039}{{\ttfamily 1901.11039}}].

\bibitem{Samtleben:2019zrh}
H.~Samtleben and {\"O}.~Sar{\i}oglu, \emph{{Consistent $S^3$ reductions of six-dimensional supergravity}}, \href{https://doi.org/10.1103/PhysRevD.100.086002}{\emph{Phys. Rev. D} {\bfseries 100} (2019) 086002} [\href{https://arxiv.org/abs/1907.08413}{{\ttfamily 1907.08413}}].

\bibitem{Cassani:2020cod}
D.~Cassani, G.~Josse, M.~Petrini and D.~Waldram, \emph{{$\mathcal{N} $ = 2 consistent truncations from wrapped M5-branes}}, \href{https://doi.org/10.1007/JHEP02(2021)232}{\emph{JHEP} {\bfseries 02} (2021) 232} [\href{https://arxiv.org/abs/2011.04775}{{\ttfamily 2011.04775}}].

\bibitem{Malek:2020jsa}
E.~Malek and V.~Vall~Camell, \emph{{Consistent truncations around half-maximal AdS$_{5}$ vacua of 11-dimensional supergravity}}, \href{https://doi.org/10.1088/1361-6382/ac566a}{\emph{Class. Quant. Grav.} {\bfseries 39} (2022) 075026} [\href{https://arxiv.org/abs/2012.15601}{{\ttfamily 2012.15601}}].

\bibitem{Josse:2021put}
G.~Josse, E.~Malek, M.~Petrini and D.~Waldram, \emph{{The higher-dimensional origin of five-dimensional $ \mathcal{N} $ = 2 gauged supergravities}}, \href{https://doi.org/10.1007/JHEP06(2022)003}{\emph{JHEP} {\bfseries 06} (2022) 003} [\href{https://arxiv.org/abs/2112.03931}{{\ttfamily 2112.03931}}].

\bibitem{Bugden:2021wxg}
M.~Bugden, O.~Hulik, F.~Valach and D.~Waldram, \emph{{G-Algebroids: A Unified Framework for Exceptional and Generalised Geometry, and Poisson{\textendash}Lie Duality}}, \href{https://doi.org/10.1002/prop.202100028}{\emph{Fortsch. Phys.} {\bfseries 69} (2021) 2100028} [\href{https://arxiv.org/abs/2103.01139}{{\ttfamily 2103.01139}}].

\bibitem{Bugden:2021nwl}
M.~Bugden, O.~Hulik, F.~Valach and D.~Waldram, \emph{{Exceptional Algebroids and Type IIB Superstrings}}, \href{https://doi.org/10.1002/prop.202100104}{\emph{Fortsch. Phys.} {\bfseries 70} (2022) 2100104} [\href{https://arxiv.org/abs/2107.00091}{{\ttfamily 2107.00091}}].

\bibitem{Galli:2022idq}
M.~Galli and E.~Malek, \emph{{Consistent truncations to 3-dimensional supergravity}}, \href{https://doi.org/10.1007/JHEP09(2022)014}{\emph{JHEP} {\bfseries 09} (2022) 014} [\href{https://arxiv.org/abs/2206.03507}{{\ttfamily 2206.03507}}].

\bibitem{Bossard:2022wvi}
G.~Bossard, F.~Ciceri, G.~Inverso and A.~Kleinschmidt, \emph{{Consistent Kaluza-Klein Truncations and Two-Dimensional Gauged Supergravity}}, \href{https://doi.org/10.1103/PhysRevLett.129.201602}{\emph{Phys. Rev. Lett.} {\bfseries 129} (2022) 201602} [\href{https://arxiv.org/abs/2209.02729}{{\ttfamily 2209.02729}}].

\bibitem{Butter:2022iza}
D.~Butter, F.~Hassler, C.N.~Pope and H.~Zhang, \emph{{Consistent truncations and dualities}}, \href{https://doi.org/10.1007/JHEP04(2023)007}{\emph{JHEP} {\bfseries 04} (2023) 007} [\href{https://arxiv.org/abs/2211.13241}{{\ttfamily 2211.13241}}].

\bibitem{Hulik:2023aks}
O.~Hulik, E.~Malek, F.~Valach and D.~Waldram, \emph{{Y-algebroids and E$_{7(7)}${\texttimes} {\ensuremath{\mathbb{R}}}$^{+}$-generalised geometry}}, \href{https://doi.org/10.1007/JHEP03(2024)034}{\emph{JHEP} {\bfseries 03} (2024) 034} [\href{https://arxiv.org/abs/2308.01130}{{\ttfamily 2308.01130}}].

\bibitem{Bossard:2023jid}
G.~Bossard, F.~Ciceri, G.~Inverso and A.~Kleinschmidt, \emph{{Consistent truncation of eleven-dimensional supergravity on S$^{8}${\texttimes} S$^{1}$}}, \href{https://doi.org/10.1007/JHEP01(2024)045}{\emph{JHEP} {\bfseries 01} (2024) 045} [\href{https://arxiv.org/abs/2309.07233}{{\ttfamily 2309.07233}}].

\bibitem{Josse:2023lsp}
G.~Josse and F.~Merenda, \emph{{Vacua scan of 5d, N=2 consistent truncations}}, \href{https://doi.org/10.1016/j.physletb.2024.138670}{\emph{Phys. Lett. B} {\bfseries 853} (2024) 138670} [\href{https://arxiv.org/abs/2310.15272}{{\ttfamily 2310.15272}}].

\bibitem{Hassler:2023axp}
F.~Hassler and Y.~Sakatani, \emph{{Hierarchy of curvatures in exceptional geometry}}, \href{https://doi.org/10.1103/PhysRevD.109.106002}{\emph{Phys. Rev. D} {\bfseries 109} (2024) 106002} [\href{https://arxiv.org/abs/2311.12095}{{\ttfamily 2311.12095}}].

\bibitem{Blair:2024ofc}
C.D.A.~Blair, M.~Pico and O.~Varela, \emph{{Infinite and finite consistent truncations on deformed generalised parallelisations}}, \href{https://doi.org/10.1007/JHEP09(2024)065}{\emph{JHEP} {\bfseries 09} (2024) 065} [\href{https://arxiv.org/abs/2407.01298}{{\ttfamily 2407.01298}}].

\bibitem{Inverso:2024xok}
G.~Inverso and D.~Rovere, \emph{{How to uplift D = 3 maximal supergravities}}, \href{https://doi.org/10.1007/JHEP02(2025)130}{\emph{JHEP} {\bfseries 02} (2025) 130} [\href{https://arxiv.org/abs/2410.14520}{{\ttfamily 2410.14520}}].

\bibitem{Guarino:2024gke}
A.~Guarino, C.~Sterckx and M.~Trigiante, \emph{{Consistent N=4, D=4 truncation of type IIB supergravity on S1{\texttimes}S5}}, \href{https://doi.org/10.1103/PhysRevD.111.046019}{\emph{Phys. Rev. D} {\bfseries 111} (2025) 046019} [\href{https://arxiv.org/abs/2410.23149}{{\ttfamily 2410.23149}}].

\bibitem{Duboeuf:2024tbd}
B.~Duboeuf, M.~Galli, E.~Malek and H.~Samtleben, \emph{{Consistent truncations and G2-invariant AdS4 solutions of D=11 supergravity}}, \href{https://doi.org/10.1103/PhysRevD.111.066007}{\emph{Phys. Rev. D} {\bfseries 111} (2025) 066007} [\href{https://arxiv.org/abs/2412.13300}{{\ttfamily 2412.13300}}].

\bibitem{Hassler:2025qhh}
F.~Hassler and Y.~Sakatani, \emph{{Consistent truncation and generalized duality based on exceptional generalized cosets}},  \href{https://arxiv.org/abs/2510.12799}{{\ttfamily 2510.12799}}.

\bibitem{Rovere:2025jks}
D.~Rovere and C.~Sterckx, \emph{{How to uplift non-maximal gauged supergravities}},  \href{https://arxiv.org/abs/2510.24850}{{\ttfamily 2510.24850}}.

\bibitem{Pico:2025cmc}
M.~Pico and O.~Varela, \emph{{Consistent subsectors of maximal supergravity and wrapped M5-branes}},  \href{https://arxiv.org/abs/2511.15892}{{\ttfamily 2511.15892}}.

\bibitem{Josse:2025uro}
G.~Josse, M.~Petrini and M.~Pico, \emph{{Consistent Truncations and Generalised Geometry: Scanning through Dimensions and Supersymmetry}},  \href{https://arxiv.org/abs/2512.03027}{{\ttfamily 2512.03027}}.

\bibitem{Pico:2026rji}
M.~Pico and O.~Varela, \emph{{Maximal trombone supergravity from wrapped M5-branes}},  \href{https://arxiv.org/abs/2601.07960}{{\ttfamily 2601.07960}}.

\bibitem{Hull:2007zu}
C.M.~Hull, \emph{{Generalised Geometry for M-Theory}}, \href{https://doi.org/10.1088/1126-6708/2007/07/079}{\emph{JHEP} {\bfseries 07} (2007) 079} [\href{https://arxiv.org/abs/hep-th/0701203}{{\ttfamily hep-th/0701203}}].

\bibitem{PiresPacheco:2008qik}
P.~Pires~Pacheco and D.~Waldram, \emph{{M-theory, exceptional generalised geometry and superpotentials}}, \href{https://doi.org/10.1088/1126-6708/2008/09/123}{\emph{JHEP} {\bfseries 09} (2008) 123} [\href{https://arxiv.org/abs/0804.1362}{{\ttfamily 0804.1362}}].

\bibitem{Coimbra:2011ky}
A.~Coimbra, C.~Strickland-Constable and D.~Waldram, \emph{{$E_{d(d)} \times \mathbb{R}^+$ generalised geometry, connections and M theory}}, \href{https://doi.org/10.1007/JHEP02(2014)054}{\emph{JHEP} {\bfseries 02} (2014) 054} [\href{https://arxiv.org/abs/1112.3989}{{\ttfamily 1112.3989}}].

\bibitem{Coimbra:2012af}
A.~Coimbra, C.~Strickland-Constable and D.~Waldram, \emph{{Supergravity as Generalised Geometry II: $E_{d(d)} \times \mathbb{R}^+$ and M theory}}, \href{https://doi.org/10.1007/JHEP03(2014)019}{\emph{JHEP} {\bfseries 03} (2014) 019} [\href{https://arxiv.org/abs/1212.1586}{{\ttfamily 1212.1586}}].

\bibitem{Hohm:2013pua}
O.~Hohm and H.~Samtleben, \emph{{Exceptional Form of D=11 Supergravity}}, \href{https://doi.org/10.1103/PhysRevLett.111.231601}{\emph{Phys. Rev. Lett.} {\bfseries 111} (2013) 231601} [\href{https://arxiv.org/abs/1308.1673}{{\ttfamily 1308.1673}}].

\bibitem{Hohm:2013vpa}
O.~Hohm and H.~Samtleben, \emph{{Exceptional Field Theory I: $E_{6(6)}$ covariant Form of M-Theory and Type IIB}}, \href{https://doi.org/10.1103/PhysRevD.89.066016}{\emph{Phys. Rev. D} {\bfseries 89} (2014) 066016} [\href{https://arxiv.org/abs/1312.0614}{{\ttfamily 1312.0614}}].

\bibitem{Hohm:2013uia}
O.~Hohm and H.~Samtleben, \emph{{Exceptional field theory. II. E$_{7(7)}$}}, \href{https://doi.org/10.1103/PhysRevD.89.066017}{\emph{Phys. Rev. D} {\bfseries 89} (2014) 066017} [\href{https://arxiv.org/abs/1312.4542}{{\ttfamily 1312.4542}}].

\bibitem{Strickland-Constable:2013xta}
C.~Strickland-Constable, \emph{{Subsectors, Dynkin Diagrams and New Generalised Geometries}}, \href{https://doi.org/10.1007/JHEP08(2017)144}{\emph{JHEP} {\bfseries 08} (2017) 144} [\href{https://arxiv.org/abs/1310.4196}{{\ttfamily 1310.4196}}].

\bibitem{Bergshoeff:2001pv}
E.~Bergshoeff, R.~Kallosh, T.~Ortin, D.~Roest and A.~Van~Proeyen, \emph{{New formulations of D = 10 supersymmetry and D8 - O8 domain walls}}, \href{https://doi.org/10.1088/0264-9381/18/17/303}{\emph{Class. Quant. Grav.} {\bfseries 18} (2001) 3359} [\href{https://arxiv.org/abs/hep-th/0103233}{{\ttfamily hep-th/0103233}}].

\bibitem{Coimbra:2011nw}
A.~Coimbra, C.~Strickland-Constable and D.~Waldram, \emph{{Supergravity as Generalised Geometry I: Type II Theories}}, \href{https://doi.org/10.1007/JHEP11(2011)091}{\emph{JHEP} {\bfseries 11} (2011) 091} [\href{https://arxiv.org/abs/1107.1733}{{\ttfamily 1107.1733}}].

\bibitem{Ashmore:2015joa}
A.~Ashmore and D.~Waldram, \emph{{Exceptional Calabi-Yau spaces: the geometry of $\mathcal{N}=2$ backgrounds with flux}}, \href{https://doi.org/10.1002/prop.201600109}{\emph{Fortsch. Phys.} {\bfseries 65} (2017) 1600109} [\href{https://arxiv.org/abs/1510.00022}{{\ttfamily 1510.00022}}].

\bibitem{Lu:1996mg}
H.~Lu, C.N.~Pope and K.S.~Stelle, \emph{{Vertical versus diagonal dimensional reduction for p-branes}}, \href{https://doi.org/10.1016/S0550-3213(96)90137-6}{\emph{Nucl. Phys. B} {\bfseries 481} (1996) 313} [\href{https://arxiv.org/abs/hep-th/9605082}{{\ttfamily hep-th/9605082}}].

\bibitem{Trigiante:2016mnt}
M.~Trigiante, \emph{{Gauged Supergravities}}, \href{https://doi.org/10.1016/j.physrep.2017.03.001}{\emph{Phys. Rept.} {\bfseries 680} (2017) 1} [\href{https://arxiv.org/abs/1609.09745}{{\ttfamily 1609.09745}}].

\bibitem{Romans:1986er}
L.J.~Romans, \emph{{Selfduality for Interacting Fields: Covariant Field Equations for Six-dimensional Chiral Supergravities}}, \href{https://doi.org/10.1016/0550-3213(86)90016-7}{\emph{Nucl. Phys. B} {\bfseries 276} (1986) 71}.

\bibitem{Riccioni:1997np}
F.~Riccioni, \emph{{Tensor multiplets in six-dimensional (2,0) supergravity}}, \href{https://doi.org/10.1016/S0370-2693(98)00070-7}{\emph{Phys. Lett. B} {\bfseries 422} (1998) 126} [\href{https://arxiv.org/abs/hep-th/9712176}{{\ttfamily hep-th/9712176}}].

\end{thebibliography}\endgroup

\end{document}